\documentclass[11pt]{article}
\usepackage{amsfonts}
\usepackage{amssymb}
\usepackage{mathrsfs}
\usepackage{graphicx}
\usepackage{amsmath,amsthm,amssymb,amscd}
\usepackage[all]{xy}
\usepackage{subfig}
\usepackage{multirow}
\usepackage{color}
\usepackage{float}
\usepackage{silence}

\newcommand{\cn}{{\cal N}}
\newcommand{\cm}{{\cal M}}

\newcommand{\cg}{{\mathfrak{g}}}

\newcommand{\co}{{\cal O}}

\newcommand{\cw}{{\cal W}}

\newcommand{\nn}{\nonumber}

\def\eqa{\begin{eqnarray}}
\def\eqae{\end{eqnarray}}
\def\eq{\begin{equation}}
\def\eqe{\end{equation}}
\def\be{\begin{equation}}
\def\ee{\end{equation}}
\def\bea{\begin{eqnarray}}
\def\eea{\end{eqnarray}}
\def\ba{\begin{array}}
\def\ea{\end{array}}
\def\bd{\begin{displaymath}}
\def\ed{\end{displaymath}}

\def\Tr{{\rm Tr}}

\def\>{\rangle}
\def\<{\langle}
\def\a{\alpha}
\def\b{\beta}

\def\del{\delta}

\def\g{\gamma}

\def\l{\lambda}
\def\m{\mu}

\def\p{\pi}

\def\r{\rho}

\def\F{\Phi}

\def\pa{\partial}

\newcommand{\bn}{{\mathbb{N}}}

\newcommand{\fn}{\footnotemark\footnotetext}

\numberwithin{equation}{section}

\begin{document}

\begin{titlepage}

\vspace{1cm}

\begin{center}
{\Large \bf The symmetry of large $\cn=4$ holography}\\[1cm]
Matthias R.\ Gaberdiel and Cheng Peng~\\[6mm]

{\it Institut f\"{u}r Theoretische Physik, ETH Zurich,}\\
{\it CH-8093 Z\"{u}rich, Switzerland}\\

\end{center}
\vspace{1cm} \centerline{\bf Abstract} \vspace*{0.5cm}

For the proposed duality relating a family of ${\cal N}=4$ superconformal  coset models to a
certain supersymmetric higher spin theory on AdS$_3$, the asymptotic symmetry algebra
of the bulk description is determined. It is shown that, depending on the choice of the boundary charges,
one may obtain either the linear or the non-linear superconformal
algebra on the boundary. We compare the non-linear version of the asymptotic
symmetry algebra with the non-linear coset algebra and find non-trivial agreement
in the 't~Hooft limit, thus giving strong support for the proposed duality. As a by-product of our analysis we also show that the ${\cal W}_\infty$
symmetry of the coset theory is broken under the exactly marginal perturbation that preserves
the ${\cal N}=4$ superconformal algebra.

\end{titlepage}

\tableofcontents

\section{Introduction}

One possible strategy for proving the AdS/CFT correspondence consists of understanding the duality
for the case when the dual conformal field theory is free (or nearly free), and then perturbing both
sides starting from this special point in moduli space. If the CFT is free it is believed that the
dual string theory
can be consistently restricted to a higher spin subsector, and thus one expects a simplified
duality between a higher spin theory on AdS and a vector-like free (or nearly free) conformal field theory
\cite{Sundborg:2000wp,Witten,Mikhailov:2002bp,Sezgin:2002rt}. Concrete proposals
for a duality of this type, relating Vasiliev higher spin theories on AdS$_4$ \cite{Vasiliev:1989re}
to ${\rm O}(N)$ vector models in $3$ dimensions were subsequently found in \cite{Klebanov:2002ja,Sezgin:2003pt},
but it was not until the work of Giombi \& Yin \cite{Giombi:2009wh,Giombi:2010vg} (see \cite{Giombi:2012ms}
for a review) when convincing non-trivial evidence in favour of these dualities was
obtained. Since then various further aspects of the duality have been understood in detail, see
in particular \cite{Maldacena:2011jn,Aharony:2011jz,Giombi:2011kc,Maldacena:2012sf}. More recently,
an embedding of this duality into string theory has also been suggested \cite{Chang:2012kt}.

In a separate development, a lower dimensional version of the duality was
suggested in \cite{Gaberdiel:2010pz}, relating a higher spin theory on AdS$_3$
\cite{Vasiliev:1995dn,Vasiliev:1999ba} to the large $N$ limit
of a family of 2d minimal model CFTs. This proposal was motivated by the
analysis of the asymptotic symmetries of higher spin theories on AdS$_3$
\cite{Henneaux:2010xg,Campoleoni:2010zq}, see also
\cite{Gaberdiel:2011wb,Campoleoni:2011hg} for subsequent developments.
This proposal has been tested and generalised in a variety of ways, for recent reviews see
\cite{GGreview,Ammon:2012wc}.

In an attempt to embed this correspondence into string theory, a large ${\cal N}=4$ version
of the duality was proposed in \cite{Gaberdiel:2013vva}; it holds the promise of relating
the higher spin theory to string theory on AdS$_3\times {\rm S}^3 \times {\rm S}^3 \times {\rm S}^1$.
This proposal was subsequently explored further. In particular, the spectrum of the two descriptions was matched
in \cite{Candu:2013fta}, see also \cite{Creutzig:2013tja} for an earlier analysis.
In this paper we add one further consistency check to this list: we establish that the asymptotic symmetry
algebra of the relevant higher spin theory matches the classical limit of the coset ${\cal W}$-algebra.
This generalises similar studies for situations with less supersymmetry
\cite{Henneaux:2010xg,Campoleoni:2010zq,Gaberdiel:2011wb,Campoleoni:2011hg,Henneaux:2012ny,Hanaki:2012yf,Peng:2012ae}.

While most of the analysis follows the well-established procedure going back to
\cite{Brown:1986nw}, there is one important subtlety that arises in the present context. The large
{${\cal N}=4$} superconformal algebra (and hence the coset theory) comes in two varieties: there is the
linear $A_\gamma$ algebra and the non-linear
$\tilde{A}_\gamma$ algebra that can be obtained from the former by quotienting out $4$ free fermions
and the $\mathfrak{u}(1)$ generator, see
\cite{Sevrin:1988ew,Schoutens:1988ig,Spindel:1988sr,VanProeyen:1989me,Sevrin:1989ce,Goddard:1988wv}
for some early literature on the subject. Naively one may expect that the dual of the higher spin theory
should lead to the non-linear version $\tilde{A}_\gamma$ of the large ${\cal N}=4$ superconformal algebra
\cite{Gaberdiel:2013vva}; however, this misses the $\mathfrak{u}(1)$ generator that corresponds to the ${\rm S}^1$ of the putative
string target space, and it is also at odds with the observation that the supergravity spectrum organises itself into representations
of the linear $A_\gamma$ algebra \cite{de Boer:1999rh}, thus suggesting that the dual CFT should also have this
symmetry. As we explain below (see section~\ref{sec:linear}) there is some freedom in the definition of the
boundary charges, and one can use this ambiguity to obtain also an asymptotic symmetry algebra from the higher spin
description that contains the linear $A_\gamma$ algebra as a subalgebra.

For the non-linear version of the algebra we then compare the structure constants of the asymptotic symmetry
algebra with those obtained from the OPEs of the ${\cal W}$-currents of the (non-linear) coset algebra in the 't~Hooft limit,
and find perfect agreement.
As a further consistency check we also study the truncation properties of both algebras. The higher spin
algebra of the AdS description can be truncated to a finite dimensional Lie algebra
for certain values of the parameters. This is inherited by the asymptotic symmetry algebra which is, for
these values, then generated by finitely many fields. We show that the dual coset algebras mirror
this truncation pattern very nicely.

The explicit description of the ${\cal W}$-currents of the coset theory also allows us to address
another question that was raised in
\cite{Gaberdiel:2013vva}: the coset theory possesses an exactly marginal field whose perturbation preserves
the large ${\cal N}=4$ superconformal algebra. However, one may suspect that this perturbation will
break the ${\cal W}_{\infty}$ symmetry, i.e., that some (or indeed all) of the higher spin currents will cease to be conserved after the
perturbation. We answer this question in section~\ref{sec:perturbation} where we show that at least
the first non-trivial higher spin current (and therefore probably all currents outside the
large ${\cal N}=4$ algebra) get broken by the perturbation.
\medskip

The paper is organised as follows. In section~2 we perform the usual asymptotic symmetry analysis
of the relevant higher spin theory that leads to the non-linear $\tilde{A}_\gamma$ subalgebra. We also
calculate some higher spin OPEs explicitly, whose details are spelled out in the appendix (as well as the
ancillary file of the {\tt arXiv} submission), and explain how the algebra can be truncated for certain values
of the parameters. In section~3 we show how the asymptotic symmetry analysis
may be adjusted so as to lead to the linear $A_\gamma$ algebra. Section~4 deals with the dual
Wolf space cosets. In particular, we determine the higher spin generators of the version of the coset that
contains the non-linear $\tilde{A}_\gamma$ superconformal symmetry, and compare the
resulting OPE coefficients to what was obtained in section~2, finding beautiful agreement.
We also explain in section~4.3 how these coset algebras can be truncated for certain values of the
parameters, again matching nicely the truncation patterns of the asymptotic symmetry algebra.
Given the explicit form of the higher spin generators of the coset theory, we finally
analyse in section~5 their behaviour under the perturbation by an exactly marginal
${\cal N}=4$ preserving perturbation. We close in section~6 with some conclusions, and
there are two appendices containing some of the detailed expressions for the OPEs, as well
as the truncation analysis for the linear $A_\gamma$ algebra.

\section{The (non-linear) asymptotic symmetry algebra}\label{sec:nonlin}

Recall that the ${\cal N}=4$ version of the higher spin/CFT duality \cite{Gaberdiel:2013vva} relates
the Wolf space coset models to the higher spin theory based on the Lie algebra shs${}_2[\l]$. In this
section we compute the asymptotic symmetry algebra of this higher spin theory, following
the basic ideas of \cite{Brown:1986nw,Campoleoni:2010zq,Henneaux:2010xg}. As we shall see, this leads to a
non-linear ${\cal W}$-algebra that contains the classical limit of
the non-linear $\tilde{A}_\gamma$ algebra as a subalgebra.
\smallskip

Let us begin by reviewing how to determine the asymptotic symmetry algebra of a general
$3$-dimensional supergravity or higher spin theory, following
\cite{Henneaux:1999ib,Campoleoni:2010zq,Henneaux:2010xg,Gaberdiel:2011wb},
see also \cite{GGreview} for a review.
We shall work with the Chern-Simons formalism that was originally introduced for the case of pure gravity in AdS$_3$
in \cite{Achucarro:1987vz,Witten:1988hc}
\begin{equation} \label{action1}
I = I_{\rm CS}(A,k_{\rm cs}) - I_{\rm CS}(\bar{A},k_{\rm cs})\ , \qquad
I_{\rm CS} (A,k_{\rm cs})= \frac{k_{\rm cs}}{4 \pi}  \int_{\cm} \mbox{Tr}(A \wedge d A + \frac{2}{3} A \wedge A \wedge A)\ .
\end{equation}
Here ${\cal M}$ is a solid cylinder, which we parametrise by $(t,\phi,\rho)$, with $\rho$ the radial direction, and
$(t,\phi)$ the parameters on the $2$-dimensional boundary cylinder.
Furthermore, $A$ and $\bar{A}$ are gauge connections in some gauge algebra $\cg$, which for the case of pure
gravity is just $\cg = \mathfrak{sl}(2,\mathbb{R})$, but in the present context will be taken to equal $\cg={\rm shs}_2[\l]$.
In order to preform the asymptotic symmetry analysis it is important that the algebra contains a preferred
$\mathfrak{sl}(2,\mathbb{R})$ subalgebra, $\mathfrak{sl}(2,\mathbb{R})\subset \cg$, whose generators we shall denote
by $L_m$ with $m=0,\pm 1$. We shall furthermore decompose the remaining generators of $\cg$ according to their
spin, i.e., we shall take a basis of $\cg$ to be given by $V^{(s)\,i}_n$, where
\be
{}[L_m, V^{(s)\,i}_{n}] = \bigl( (s-1)m - n \bigr)\, V^{(s)\,i}_{m+n} \ .
\ee
(Here $i$ is some additional index that labels the different generators of the same spin $s$.)
The choice of an $\mathfrak{sl}(2,\mathbb{R})$ subalgebra allows us to define what we mean by the AdS$_3$ vacuum solution,
\be
A_{\rm AdS} = b^{-1} \Bigl(L_1 + \frac{1}{4} L_{-1} \Bigr) \, b\, \,dx^+ +
b^{-1} \partial_\rho b \, d\rho \label{AAdS}
\ee
with a similar expression for $\bar{A}$, where
\be
b(\rho) = e^{\rho L_0} \ .
\ee
We are interested in solutions that are asymptotically AdS, i.e., we shall consider gauge connections $A$ that satisfy
for $\rho\rightarrow \infty$
\be\label{asads}
A-A_{\rm AdS}\sim \co (1)\ .
\ee
Then, following the analysis in \cite{Campoleoni:2010zq}, we can choose a gauge so that
\be\label{gxcon}
A=b(\r)^{-1} a(x^+) b(\r)\ , \qquad b(\r)=e^{\r L_0}\ , \qquad
a(x^+)= L_{1}+\sum_{s,i} a^{(s)\,i} \, V^{(s)\,i}_{1-s} \ .
\ee
The asymptotic symmetry algebra is now the residual gauge symmetry that leaves the form of the gauge fixed connection
\eqref{gxcon} unchanged. More concretely, under an arbitrary gauge transformation labelled by $\gamma\in\cg$, the
gauge connection changes as
\be
\del_\g a=\sum\limits_{s,i}c^{(s)\,i}_m V^{(s)\,i}_m=d\g+[a,\g]\ ,
\ee
where the $c^{(s)\,i}_m$ depend in general on $a$ (as well as $\gamma$).
Then requiring the gauge connection to preserve the AdS boundary condition and to stay in the gauge (\ref{gxcon})
implies that
\be\label{cond}
c^{(s)\, i}_m=0\ ,\qquad  m\neq 1-s\ ,\quad \forall s,i\ .
\ee
This leads to a set of equations for the gauge parameters that one can solve recursively, starting with any choice for the
`highest' component  $\gamma^{(s)\, i}_{s-1}$ of $\gamma$.  The resulting gauge symmetries then define the asymptotic
symmetries of the AdS theory.
\smallskip

In order to endow this set of asymptotic symmetries with the structure of a Poisson algebra, we now associate to each
gauge transformation a conserved charge as
\be\label{charge}
Q(\gamma) = - \frac{k_{\rm cs} }{2\pi} \int {\rm Tr}( \gamma a) \ ,
\ee
where the integral is taken along $\phi$ in the boundary cylinder of AdS$_3$.  Then we define the Poisson bracket via
\be\label{poisson}
\{ Q(\gamma),X \} = \delta_\gamma X \ ,
\ee
where the right-hand side is the gauge variation of $X$ under the gauge transformation described by $\gamma$
(where $\gamma$ satisfies \eqref{cond}); this analysis is explained in more detail in \cite{Campoleoni:2010zq},
see also \cite{GGreview} for a review.
The resulting Poisson algebra should then be identified with the classical limit of the ${\cal W}$-algebra
underlying the dual conformal field theory.

\subsection{The asymptotic symmetry algebra of the higher spin theory}

In this section we describe the results of this analysis for the case of $\cg = {\rm shs}_2[\l]$ (with the choice of the
$\mathfrak{sl}(2,\mathbb{R})$ subalgebra as described in \cite{Gaberdiel:2013vva}). We have found that the asymptotic
symmetry algebra one obtains in this manner is a non-linear ${\cal W}$-algebra. It contains, in particular, the non-linear
$\tilde{A}_\gamma$ ${\cal N}=4$ superconformal algebra as was already anticipated in \cite{Gaberdiel:2013vva}. Indeed,
this algebra is the natural asymptotic symmetry algebra associated to the subalgebra $D(2,1|\alpha)\subset {\rm shs}_2[\l]$.

More concretely, $D(2,1|\alpha)$ consists of an $\mathfrak{sl}(2,\mathbb{R})$ subalgebra generated by the wedge
modes of the stress energy tensor $T$, the wedge modes of the $4$ supercharges that we denote by $G^{\pm\pm}$, as well
as the wedge(=zero) modes of $6$ currents transforming
in the Lie algebra $\mathfrak{su}(2) \oplus \mathfrak{su}(2)$; we shall  denote these currents as $A^{+ a}$ and $A^{-a}$, respectively,
where $a=\pm, 3$ denotes the components of $\mathfrak{su}(2)$ in the Cartan-Weyl basis. The asymptotic symmetry
algebra is then generated by the associated fields, and their Poisson brackets turn out to equal
(in the following we have translated the Poisson brackets into `classical' OPEs, i.e., OPEs without any
normal ordering of composite fields, and have only written the singular terms)
\begin{eqnarray}
G^{++}(z)\, G^{++}(w) &\sim& -\frac{8 (\gamma -1) \gamma  A^{-+}(w)\,A^{++}(w)}{k_{\rm cs} (w-z)} \label{GppGpp}\\
G^{++}(z)\, G^{+-}(w) &\sim&\frac{8 (\gamma -1) \gamma  A^{-3}(w)\,A^{++}(w)}{k_{\rm cs} (w-z)}
-\frac{4 \gamma  \pa A^{++}(w)}{w-z}+\frac{8 \gamma  A^{++}(w)}{(w-z)^2} \label{GppGpm}\\
G^{++}(z)\, G^{-+}(w) &\sim& \frac{8 (\gamma -1) \gamma  A^{-+}(w)\,A^{+3}(w)}{k_{\rm cs} (w-z)}
+\frac{4 (\gamma -1) \pa A^{-+}(w)}{w-z} \nn \\
& &   -\frac{8 (\gamma -1) A^{-+}(w)}{(w-z)^2}\label{GppGmp}
\end{eqnarray}
\begin{eqnarray}
\nn  G^{++}(z)\, G^{--}(w) &\sim&  \frac{2 (\gamma -1) \gamma  A^{--}(w)\,A^{-+}(w)}{k_{\rm cs} (w-z)}
+\frac{2 (\gamma -1) \gamma  A^{-+}(w)\,A^{--}(w)}{k_{\rm cs} (w-z)}\nonumber \\
& & +\frac{4 (\gamma -1) \gamma  A^{-3}(w)\,A^{-3}(w)}{k_{\rm cs} (w-z)}
+\frac{2 (\gamma -1) \gamma  A^{+-}(w)\,A^{++}(w)}{k_{\rm cs} (w-z)}
\nonumber \\
& & +\frac{2 (\gamma -1) \gamma  A^{++}(w)\,A^{+-}(w)}{k_{\rm cs} (w-z)}
+\frac{4 (\gamma -1) \gamma  A^{+3}(w)\,A^{+3}(w)}{k_{\rm cs} (w-z)}
\nn \\
&&-\frac{8 (\gamma -1) \gamma  A^{-3}(w)\,A^{+3}(w)}{k_{\rm cs} (w-z)} \nn \\
& & +\frac{4 \big(-\gamma  \pa A^{-3}(w)+\gamma  \pa A^{+3}(w)+\pa A^{-3}(w)+T(w)\big)}{w-z} \nn \\
&&+\frac{8 (\gamma  A^{-3}(w)-\gamma  A^{+3}(w)-A^{-3}(w))}{(w-z)^2}+\frac{8 k_{\rm cs} }{(w-z)^3}\ , \label{GppGmm}
\end{eqnarray}
with similar expressions for the remaining generators.
These OPEs agree precisely with those of the non-linear $\tilde{A}_\gamma$ algebra, as given for example in
appendix~B.3 of \cite{Gaberdiel:2013vva}, provided we identify
\be\label{krel}
k_{\rm cs}=\frac{k^+k^-}{k^++k^-}\ , \qquad \hbox{and} \qquad
\g \equiv \lambda = \frac{k^-}{k^++k^-} \ ,
\ee
and drop subleading (i.e., $1/k^\pm$) corrections. Note that (\ref{krel}) is the
expected relation, given the comparison of the central charges.

Actually, we can test the correspondence between the symmetry algebras further. As is explained in
\cite{Gaberdiel:2013vva}, the algebra ${\rm shs}_2[\l]$ contains in addition to $D(2,1|\alpha)$ other
generators that organise themselves into supermultiplets of $D(2,1|\alpha)$. The lowest non-trivial
multiplet (denoted by $R^{(1)}$ in that paper) is generated from a Virasoro primary operator of spin $1$
that transforms in a singlet representation under $\mathfrak{su}(2)\oplus \mathfrak{su}(2)$. We shall
denote it by $V^{(1)0}$ in the following, and introduce for the superdescendants the notation

\begin{center}
\begin{tabular}{|c|c|c|c|}
  \hline
  fields & $h$ & $(l^+,l^-)$ & multiplet\\\hline
$V^{(1)0}$ & 1 & $(0,0)$ & $R^{(1)}$ \\\hline
$G'^{\pm\pm}$ & $\frac{3}{2}$ &$(\frac{1}{2},\frac{1}{2})$  & $R^{(1)}$\\\hline
 $V^{(2)++}\,,V^{(2)+-}\,,V^{(2)+3}$ & 2 & $(1,0)$ & $R^{(1)}$ \\\hline
  $V^{(2)-+}\,,V^{(2)--}\,,V^{(2)-3}$ & 2 & $(0,1)$& $R^{(1)}$\\
  \hline
\end{tabular}
\end{center}

\noindent
The other generator we shall consider in the following is the primary component of spin $2$ in the second
multiplet $R^{(2)}$, which we shall denote by $V^{(2) 0}$. We have  evaluated some of the
Poisson brackets of these generators, and the explicit expressions are given (in OPE language)
in appendix~\ref{opeapd} --- additional OPEs are also listed in the ancillary file of the {\tt arXiv} submission.
In section~\ref{cftsec} below we shall calculate these OPEs also from the dual coset CFT viewpoint, and show
that the above structure constants match with the CFT answer in the 't~Hooft limit. In addition, we shall also
explain there that the truncation patterns of the asymptotic symmetry algebra  (to which we now turn) are
nicely reproduced by the dual CFT.

\subsection{Truncation of the asymptotic symmetry algebra}\label{astruncation}

As pointed out in \cite{Gaberdiel:2013vva}, the higher spin algebra ${\rm shs}_2[\mu]$ admits a truncation
at $\mu=\g=s+1$ with $s\in\bn$. For this value of $\mu$, the higher spin algebra contains a
large ideal, and the quotient by it is the finite dimensional Lie algebra
\begin{equation}
{\rm shs}_2[s+1] = D(2,1| - \tfrac{s+1}{s}) \oplus \bigoplus_{i=1}^{s-1} R^{(i)} \ \oplus \hat{R}^{(s)}_- \ ,
\end{equation}
where $\hat{R}^{(s)}_-$ is the short representation of $D(2,1| \alpha)$ with $\alpha = - \tfrac{s+1}{s}$ whose spin content
is   \cite{Gaberdiel:2013vva}
\be
\begin{array}{lll}\label{D2multshort}
& s: & \quad ({\bf 1},{\bf 1})  \\
\hat{R}^{(s)}_-: \qquad  & s+\tfrac{1}{2}: & \quad ({\bf 2},{\bf 2})  \\[2pt]
& s+1: & \quad ({\bf 1},{\bf 3})  \ .
\end{array}
\ee
There is a similar truncation that happens at $\mu=\g=-s$ with $s\in\bn$,
where at level $s+1$ the representation $({\bf 3},{\bf 1})$ (instead of
$({\bf 1},{\bf 3})$) is retained. The fact that there are two truncations is a consequence of the
$\a \leftrightarrow \a^{-1} $ isomorphism of the $D(2,1|\a)$ algebra (see \cite{Gaberdiel:2013vva}
for more details).

It is then natural to ask whether the associated asymptotic symmetry algebra is similarly truncated, i.e., whether
the fields associated to the modes that lie in the ideal  of ${\rm shs}_2[\mu]$ also form an ideal in the asymptotic symmetry
algebra. With the explicit calculations we have done, we can test this for the simplest truncation that occurs for
$\gamma=\mu=2$ (or $\gamma=\mu=-1$), for which the truncated higher spin algebra consists of $D(2,1|\alpha)$ (with $\alpha=-2$
or $\alpha=-\frac{1}{2}$, respectively), together with one
of the two possible shortened $R^{(1)}$ representations, respectively.

Using the explicit results that are described in the ancillary file, we have checked that the fields from the multiplet
$V^{(2)}$ form indeed an ideal; for example, the $GG$ bilinear terms that appear in the OPE
$V^{(2)-+} \, V^{(2)0}$ disappear at $\gamma=-1$ (since their coefficient is proportional to $(\gamma+1)$).
For $\gamma=2$ the analysis works similarly, the only subtlety being that the OPE $G^{++}\, V^{(2)0}$ becomes
singular at $\gamma=2$. This is an artificial singulartiy, reflecting the fact that we have not normalised  the field
$V^{(2)0}$ correctly. Once this is taken into account (by rescaling $V^{(2)0} \mapsto (\gamma-2)\, V^{(2)0}$)
the $GG$ bilinear terms in this OPE drop out at $\gamma=2$.

\section{Obtaining the linear large ${\cal N}=4$ algebra from the bulk}\label{sec:linear}

As was already mentioned in \cite{Gaberdiel:2013vva}, $D(2,1|\alpha)$ is not only the `wedge' algebra of
the non-linear ${\cal N}=4$ superconformal algebra $\tilde{A}_\gamma$, but also of the
linear ${\cal N}=4$ superconformal algebra $A_\gamma$. (For some early literature on the
two ${\cal N}=4$ algebras and their relation, see
\cite{Sevrin:1988ew,Schoutens:1988ig,Spindel:1988sr,VanProeyen:1989me,Sevrin:1989ce,Goddard:1988wv}.
We follow here the notation of \cite{Gaberdiel:2013vva}, see in particular appendix~B of that paper.)
In the previous section we determined the asymptotic symmetry algebra  of the ${\rm shs}_2[\l]$ higher spin
theory and saw that it is a ${\cal W}$-algebra that contains the non-linear superconformal
algebra $\tilde{A}_\gamma$  as a subalgebra. However, one may ask whether this is necessarily so, or whether
it is also possible to obtain instead an asymptotic symmetry algebra that contains
the linear $A_\gamma$ algebra as a subalgebra. In the following
we want to explain that the second possibility also arises provided one modifies the definition of the charges
by a suitable boundary term.
\smallskip

Before going into the detailed constructions, let us first summarise our strategy. One important difference between the linear
$A_\g$ algebra and the nonlinear $\tilde{A}_\g$ algebra is the presence of spin-$\frac{1}{2}$ fields. However, these fields are not
captured by the bulk theory since the ${\rm shs}_2[\l]$ algebra does not contain any spin-$\frac{1}{2}$ generators. Our main task is
therefore to introduce auxiliary fields (that are not in the ${\rm shs}_2[\l]$ algebra) into the asymptotic symmetry algebra.
These fields will make their appearance in various correction terms which we shall add to the conserved charges \eqref{charge}. In
turn, these correction terms will then also modify the Poisson brackets via (\ref{poisson}), and for a judicious choice of these
auxiliary fields, we can recover the linear $A_\gamma$ subalgebra.

\subsection{Modification of the global charge}

Recall that the Poisson bracket of the asymptotic symmetry algebra was determined from the variation
of the associated charges, see eqs.~(\ref{charge}) -- (\ref{poisson}). We now want to
modify the conserved charges as\fn{This is a generalisation of
\cite{Benguria:1976in} where the global charge $Q(\g)$ was introduced.}
\be\label{newq}
{Q'}(\gamma)= -\frac{k_{\rm cs}}{2\p} \int \Tr(\gamma\, a')\ , \qquad a'(x^+)= L_{1}+\sum_{s,i} a'^{(s)\,i}\, V^{(s)\,i}_{1-s} \ ,
\quad a'^{(s)\,i}=a^{(s)\,i}+b^{(s)\, i} \ ,
\ee
where $b^{(s)\, i}$ are suitable auxiliary boundary fields that we will introduce in the following. (As it will turn out,
the $b^{(s)\, i}$ will actually be polynomials in some set of auxiliary fields that we shall add.)
These auxiliary boundary fields will be independent of the gauge connection $a$, and hence the gauge variation, i.e., $\delta_\gamma b=0$
for any gauge variation $\gamma$. Here $\gamma$ refers to an `unmodified' gauge variation, i.e., one that involves only
the original $a$-fields. We obviously need to postulate how the auxiliary fields transform under gauge transformations involving
the $b$-fields themselves. Since we do not want to add non-trivial degrees of freedom on the boundary, we shall take these
gauge variations to be those of free fields, i.e., we shall take the auxiliary fields we add to have the OPEs of free fields.

\noindent Then the Poisson brackets of the modified charges take the form
\be\label{algaftermod}
\{Q'(\gamma_1),Q'(\gamma_2)\}=Q'([\gamma_1,\gamma_2])
+\frac{k_{\rm cs}}{2\p} \int \Tr (\gamma_1\,d\gamma_2+b\,[\gamma_1,\gamma_2] - \gamma_2 \,\delta'_{1}b)\ ,
\ee
where the last term describes the variation due to the modified gauge variation corresponding to $\gamma_1$.
Since the new `gauge' variations only differ from the original ones by OPEs of free fields that are gauge singlets
with respect to the old gauge variations,  the modified Poisson brackets will
still satisfy the Jacobi identities.

As we shall see in the following, by introducing suitable auxiliary boundary degrees we can
remove some of the non-linearities of the asymptotic symmetry algebra; in particular, we can linearise the
non-linear $\tilde{A}_\gamma$ algebra leading to the linear $A_\gamma$ algebra. (In effect, this is just reversing
the process by means of which the free fermions and the $\mathfrak{u}(1)$-field are removed in going
from the linear to the non-linear algebra \cite{Goddard:1988wv}.) We have also attempted to
linearise the entire asymptotic symmetry algebra, but as far as we can make out, this will not be possible
by a mechanism of this sort.

\subsection{Linearising $\tilde{A}_\gamma$}

The auxiliary fields we shall now add are $4$ spin-$\frac{1}{2}$ fields $f^{\pm\pm}$ as well as a field $u$ of
conformal dimension one. These fields themselves will not appear in the modified charges, but
certain polynomials of them will, i.e., we will set
\be
b^{(s)\, i}=b^{(s)\, i}(f^{\pm\pm},u)\,,
\ee
where $b^{(s)\, i}(f^{\pm\pm},u)$ are polynomials whose degree is determined by the spin $s$. Initially, these fields
will have trivial OPEs with the fields that are originally present in the asymptotic symmetry algebra, and we postulate
their OPEs among themselves to take the form
\be
u(z)\,u(w)\sim\frac{c_u}{(z-w)^2}\ ,
\ee
where $c_u$ is some normalisation constant, as well as
\be
f^{++}(z)\,f^{--}(w)\sim\frac{4}{(z-w)}\ , \qquad f^{+-}(z)\,f^{-+}(w)\sim-\frac{4}{(z-w)} \ .
\ee
Then we can calculate the new Poisson brackets following (\ref{algaftermod}). We still have the freedom to choose the
polynomials $b^{(s)\, i}$ of the auxiliary fields, and we want to use this freedom to linearise the OPEs of the low-lying
fields. As it turns out, this can be achieved provided we proceed as follows. First we add polynomial terms to the spin $1$-fields as
\bea
A'^{++} & =A^{++}-\frac{1}{4}\, f^{++}\,f^{+-} \ , \qquad A'^{+-} &=A^{+-}-\tfrac{1}{4}\, f^{--}\,f^{-+}\ ,  \nn \\
A'^{-+} & = A^{-+}-\frac{1}{4} \, f^{++}\,f^{-+}\ , \qquad  A'^{--}&= A^{--}-\tfrac{1}{4} \,f^{--}\,f^{+-}\ ,
\eea
as well as
\bea
A'^{+3}& = & A^{+3}+ \tfrac{1}{8} ( f^{-+}\,f^{+-}+ f^{++}\,f^{--}) \nn \\
A'^{-3} &= & A^{-3}+\tfrac{1}{8} (- f^{+-}\,f^{-+}+ f^{++}\,f^{--})\ .\label{modA}
\eea
This redefinition guarantees, in particular, that the free fermion fields $f^{\pm\pm}$ transform then in the
$({\bf 2},{\bf 2})$ with respect to the gauge fields $A'$. Similarly, we modify the stress-energy tensor so that
the auxiliary fields obtain the correct conformal dimension
\bea
&&T'=T+\frac{1-\g}{k_{\rm cs}} \bigl(A'^{-3}A'^{-3}+\tfrac{1}{2}(A'^{--}A'^{-+}+A'^{-+}A'^{--}) \bigr)\\
\nn&&\qquad +\frac{\g}{k_{\rm cs}} \bigl(A'^{+3}A'^{+3}+\tfrac{1}{2}(A'^{+-}A'^{++}+A'^{++}A'^{+-})\bigr)+\frac{1}{2c_u} u^2 \\
\nn&&\qquad +\frac{1}{8}\big((\pa f^{--})f^{++}+(\pa f^{++})f^{--}-(\pa f^{+-})f^{-+}-(\pa f^{-+})f^{+-}\big)\\[2pt]
\nn&&\qquad -\frac{1-\g}{4 k_{\rm cs}}\big(-f^{--}f^{+-}A'^{-+}+f^{+-}f^{-+}A'^{-3}+f^{++}f^{--}A'^{-3}-f^{++}f^{+-}A'^{-+}\big)\\[2pt]
\nn&& \qquad -\frac{\g}{4 k_{\rm cs}}\big(-f^{--}f^{-+}A'^{++}+f^{-+}f^{+-}A'^{+3}+f^{++}f^{--}A'^{+3}+f^{++}f^{+-}A'^{+-}\big)\\[2pt]
&& \qquad -\frac{3(2 \g-1)}{32 k_{\rm cs}}f^{--}f^{++}f^{-+}f^{+-}\,.\label{modT}
\eea
The modified stress-energy tensor leads to a (classical) Virasoro algebra with central charge
$c=6k_{\rm cs}$. Finally the spin-$\frac{3}{2}$ fields  are modified as
\bea
\nn G'^{++}&=&G^{++} + \frac{i \, u\,f^{++}}{ \sqrt{2c_u}} -\frac{\sqrt{-(\gamma -1) \gamma }\,
\left(A^{-+}\,f^{+-}+A^{-3}\,f^{++}\right)}{\sqrt{k_{\rm cs}}}\\
\nn &&\qquad +\frac{\sqrt{-(\gamma -1) \gamma }\,  \left(A^{++}\,f^{-+}+A^{+3}\,f^{++}\right)}{\sqrt{k_{\rm cs}}}
+ \frac{\sqrt{-(\gamma -1) \gamma } \, f^{++}\,f^{+-}\,f^{-+}}{2\sqrt{k_{\rm cs}}} \\[6pt]
\nn G'^{-+}&=&G^{-+} + \frac{ i\,  u\,f^{-+}}{\sqrt{2 c_u}}
-\frac{ \sqrt{-(\gamma -1) \gamma }\,  \left(A^{-+}\,f^{--}+A^{-3}\,f^{-+}\right)}{\sqrt{k_{\rm cs}}}\\
\nn &&\qquad +\frac{\sqrt{-(\gamma -1) \gamma }\,
\left(A^{+-}\,f^{++}-A^{+3}\,f^{-+}\right)}{\sqrt{k_{\rm cs}}}
+ \frac{\sqrt{-(\gamma -1) \gamma }\, f^{++}\,f^{--}\,f^{-+}}{2\sqrt{k_{\rm cs}}} 
\eea
\bea
\nn G'^{+-}&=&G^{+-} +\frac{ i\, u\,f^{+-}}{\sqrt{2c_u}}
-\frac{ \sqrt{-(\gamma -1) \gamma }\, \left(A^{--}\,f^{++}-A^{-3}\,f^{+-}\right)}{\sqrt{k_{\rm cs}}}\\
\nn &&\qquad +\frac{ \sqrt{-(\gamma -1) \gamma }\,
\left(A^{++}\,f^{--}+A^{+3}\,f^{+-}\right)}{\sqrt{k_{\rm cs}}}
+\frac{\sqrt{-(\gamma -1) \gamma }\, f^{++}\,f^{+-}\,f^{--}}{2 \sqrt{k_{\rm cs}}} \\[6pt]
\nn G'^{--}&=&G^{--} +\frac{ i\, u\, f^{--}}{\sqrt{2c_u}}
- \frac{\sqrt{-(\gamma -1) \gamma } \, \left(A^{--}\,f^{-+}-A^{-3}\,f^{--}\right)}{\sqrt{k_{\rm cs}}}\\
\nn&&\qquad +\frac{\sqrt{-(\gamma -1) \gamma }\, \left(A^{+-}\,f^{+-}-A^{+3}\,f^{--}\right)}{\sqrt{k_{\rm cs}}}
-\frac{\sqrt{-(\gamma -1) \gamma } \, f^{-+}\,f^{--}\,f^{+-}}{2 \sqrt{k_{\rm cs}}} \ . \label{modG}
\eea
We have checked that these fields (together with the $f^{\pm\pm}$ and the $u$-fields) then generate
the OPEs of the linear $A_\gamma$ algebra as given in mode form for example
in \cite[appendix~B.1]{Gaberdiel:2013vva} in the 't~Hooft limit;
in order to obtain an exact match we need to rescale the fields as
\be
f^{**}=
f'^{**}\sqrt{\frac{4\g(1-\g)}{k_{\rm cs}}}\ ,\qquad \text{and}\qquad u=u' \sqrt{\frac{2 c_u\g(1-\g)}{k_{\rm cs}}}\ .
\ee
Furthermore, the parameters are identified as
\be
k_{\rm cs}=\frac{k^+k^-}{k^++k^-}\ ,\qquad \text{and}\qquad  \g=\frac{k^-}{k^++k^-}\ .
\ee
Given that the algebra is linear, one may have expected that this match also works at finite $k^\pm$, i.e., without
taking the 't~Hooft limit. However, since the redefinition of the generators involves non-linear terms,
there are normal-ordering contributions that are missed in the classical analysis, and thus we need to take the
't~Hooft limit of the quantum algebra  in order to see agreement.

\section{The dual coset CFT}\label{cftsec}

After this interlude we now come back to the main subject of the paper: we want to relate the asymptotic symmetry
algebra of the higher spin theory (that was determined in section~\ref{sec:nonlin})
to the chiral algebra of the dual coset CFT
\be\label{scoset}
\Bigl[ \frac{\mathfrak{su}(N+2)^{(1)}_{k+N+2}}{\mathfrak{su}(N)^{(1)}_{k+N+2} \oplus \mathfrak{u}^{(1)}(1)_\kappa} \Bigr] / \bigl(
\hbox{$3$ free fermions} \bigr) \ ,
\ee
where $\mathfrak{g}^{(1)}$ denotes the ${\cal N}=1$ superconformal affine algebra associated to $\mathfrak{g}$,
and we have divided out the $3+1$ free fermions and the $\mathfrak{u}(1)$ current of the linear $A_\gamma$ subalgebra
in order to obtain the non-linear $\tilde{A}_\gamma$ algebra.\footnote{The fourth fermion is part of the
$\mathfrak{u}(1)^{(1)}$ algebra. We should also mention that, for the case $N=3$, the version of the coset
algebra that contains the linear $A_\gamma$ algebra was  studied in some detail in \cite{Ahn:2013oya}.}
In bosonic language we can therefore describe the coset as
\be\label{boscos}
\frac{ \mathfrak{su}(N+2)_{k} \oplus \mathfrak{so}(4N)_1}{\mathfrak{su}(N)_{k+2} \oplus \mathfrak{u}(1)_\kappa} \ ,
\ee
where $\mathfrak{so}(4N)_1$ describes the $4N$ free fermions that survive.
The basic structure of this coset algebra was already explained in \cite{Gaberdiel:2013vva}, see
in particular section~3 of that paper; we shall essentially follow the same conventions as there,
but want to be somewhat more explicit so as to be able to calculate some of the structure constants of the
algebra. (For some of these computations we have used the Mathematica package
\texttt{OPEdefs.m} of Thielemans \cite{Thielemans:1991uw}.)
These results can then be compared with what we found in section~\ref{sec:nonlin} in the 't~Hooft limit,
i.e., in the limit $N,k\rightarrow\infty$ with
\be\label{thooft}
\frac{N}{N+k} = \lambda \quad \hbox{fixed.}
\ee
\smallskip

Let us denote by ${\cal J}^A$ the currents of the numerator $\mathfrak{su}(N+2)^{(1)}_{k+N+2}$ algebra, and
by$J^A$ the currents of the associated bosonic subalgebra $\mathfrak{su}(N+2)_{k}$ that commute with
the free fermions in the adjoint representation. Under the decomposition
$\mathfrak{su}(N+2) \supset \mathfrak{su}(N) \oplus \mathfrak{su}(2)$, the adjoint representation of
$\mathfrak{su}(N+2)$ decomposes as
\be\label{addec}
\mathfrak{su}(N+2) = \mathfrak{su}(N) \oplus \mathfrak{su}(2) \oplus \mathfrak{u}(1)
\oplus ({\bf N},{\bf 2}) \oplus (\bar{\bf  N},{\bf 2}) \ .
\ee
We denote the subset of the bosonic $\mathfrak{su}(N+2)_k$ currents that transform in the $({\bf N},{\bf 2})$ and $(\bar{\bf  N},{\bf 2})$
as
\be\label{bosfund}
J^{i,\a}=(t^A_{\Box})_{\a+N,i}\, J^{A}\ ,
\qquad \bar{J}^{i,\a}=(t^A_{\bar\Box})_{\a+N,i}\, J^{A}=-(t^A_{\Box})_{i,\a+N}\, J^{A}\ ,\quad \a=1,2 \ .
\ee
Here the adjoint index $A$ is implicitly summed over, and $t^A_\Box$ are the (traceless) $(N+2)\times (N+2)$ matrices in the
fundamental representation of $\mathfrak{su}(N+2)$. The index $i$ takes the values $i\in\{1,\ldots,N\}$, while
$\alpha=1,2$ labels the two states in the ${\bf 2}$ of $\mathfrak{su}(2)$. So the generators $J^{i,\a}$ are simply
the generators of the adjoint representation that correspond to the first $N$ entries of the last
two rows in the fundamental representation, while $\bar{J}^{i,\a}$ are those that correspond to minus the
first $N$ entries in the last two columns. We shall order the generators $J^A$
so that the first $N^2-1$ generators are those that describe the  generators of the $\mathfrak{su}(N)_k$ subalgebra
(that correspond to the upper $N\times N$ block in the fundamental representation)
\be\label{Ja}
J^a=J^A\ ,
 \qquad a\equiv A\leq N^2-1\ .
\ee
(We shall also adopt the same ordering convention for the fermions.)
Then the singular parts of the OPEs with $J^{i,\a}$ and $\bar{J}^{i,\a}$ take the form
\be\label{suNtrans}
J^a(z) \, J^{i,\a}(w) \sim \frac{J^{j,\a}\, (t^a_{\Box})_{j,i}}{(z-w)}\ , \quad \hbox{and} \quad
J^a(z) \, \bar{J}^{i,\a}(w) \sim \frac{\bar{J}^{j,\a}\, (t^a_{\bar\Box})_{j,i}}{(z-w)}\ ,
\ee
while the singular part of the OPE of $J^{i,\a}$ with $\bar{J}^{j,\b}$ equals
\be
J^{i,\a}(z)\, \bar{J}^{j,\b}(w) \sim
\frac{ - k\del_{ij}\del_{\a\b} }{(z-w)^2}-\frac{ \bigl(\del_{\a\b}(t^C_\Box)_{j,i}-(t^C_\Box)_{\a+N,\b+N}\del_{ij}\bigr)\, J^{C}(w)}{(z-w)} \ ,
\ee
and
\be
\bar{J}^{i,\a}(z)\, {J}^{j,\b}(w)
=\frac{ - k\del_{ij}\del_{\a\b} }{(z-w)^2}+\frac{ \bigl(\del_{\a\b}(t^C_\Box)_{i,j}-(t^C_\Box)_{\b+N,\a+N}\del_{ij}\bigr) J^{C}(w)}{(z-w)}\ .
\ee
Here we are working with hermitian representation matrices for which $t^A_\Box$ only has one pair of non-zero entries,
and we have chosen the normalisation convention
\be
{\rm Tr}(t^A_\Box \, t^B_\Box)= \del^{AB} \ ,
\ee
so that for all $\alpha,\beta\in\{1,2\}$ and $i,j\in\{1,\ldots,N\}$
\be \label{onml}
\sum\limits_A(t^A_\Box)_{\a+N,i}\, (t^A_{\Box})_{j,\b+N}  = \del_{ij}\, \del_{\a\b}\ , \qquad
\sum\limits_A(t^A_\Box)_{i,\a+N}\, (t^A_{\Box})_{j,\b+N} =  0\ ,
\ee
as well as
\be
\label{onmln2}
\sum\limits_A(t^A_\Box)_{rs} \, (t^A_{\Box})_{uv} =
\Big( \del^{rv}\del^{su}- \frac{\del^{rs}\del^{uv}}{(N+2)}\Big)\ ,
\ee
where $r,s,u,v\in\{1,\ldots,N+2\}$.

\subsection{The description of the fermions}

The surviving fermionic fields $\psi^{i,\a}$ and $\bar\psi^{i,\a}$ correspond to the
representations $({\bf N},{\bf 2})$ and $(\bar{\bf N},{\bf 2})$ in (\ref{addec}) above; we denote them
analogously to the bosons as
\be\label{fermfund}
\psi^{i,\a}=(t^A_\Box)_{\a+N,i} \, \psi^A\ ,\qquad
\bar\psi^{i,\a}=(t^A_{\bar{\Box}})_{\a+N,i} \, \psi^A=-(t^A_{\Box})_{i,\a+N} \, \psi^A\,.
\ee
They satisfy the standard free field OPEs
\be
\psi^{i,\a}(z)\, \bar\psi^{j,\b}(w) \sim -\frac{\del_{ij}\del_{\a\b}}{(z-w)}\ , \qquad
\bar\psi^{i,\a}(z)\, \psi^{j,\b}(w) \sim
-\frac{\del_{ij}\del_{\a\b}}{(z-w)}\ ,
\ee
as well as
\be
\psi^{i,\a}(z)\, \psi^{j,\b}(w) \sim {\cal O}(1) \ , \qquad
\bar\psi^{i,\a}(z)\, \bar\psi^{j,\b}(w) \sim {\cal O}(1) \ .
\ee
They also transform in the (anti-)fundamental representation of the
$\mathfrak{su}(N)_{k+2}$ algebra in the denominator of (\ref{boscos}); note that this
algebra is obtained from the $\mathfrak{su}(N)_{k}$ algebra generated by the $J^a$ (see eq.~(\ref{Ja}) above)
by adding bilinear fermionic  terms that are schematically of the form
\be\label{cosetcur}
J^a - (t^a_{\Box})_{ij}\psi^{i\beta} \bar\psi^{j\beta} \ .
\ee
The above free fermions also give rise to an $\mathfrak{su}(2)_N$ affine algebra
(that commutes with $\mathfrak{su}(N)_{k+2}$)
\be
K^{\a\b}=:\psi^{i,\a}\bar\psi^{i,\b}:\ ,
\ee
with respect to which the fermions transform as
\be
K^{\a\b}(z)\, \psi^{i,\g}(w) \sim -\frac{\psi^{i,\a}\del^{\b\g}}{z-w}\ , \qquad
K^{\a\b}(z)\, \bar\psi^{i,\g}(w) \sim \frac{\bar\psi^{i,\b}\del^{\a\g} }{z-w}\ .
\ee
The other $\mathfrak{su}(2)_k$ algebra that appears in $\tilde{A}_\gamma$ can simply be identified with the currents
\be
J^{\a\b}=(t^A_\Box)_{\a+N,\b+N} \, J^{A}\ ,
\ee
i.e., with the components of $J^A$ corresponding to the bottom $2\times 2$ block. These generators also
commute with the $\mathfrak{su}(N)_{k+2}$ algebra from (\ref{cosetcur}), and they define an
$\mathfrak{su}(2)$
algebra at level $k$.
In fact, in both cases we only get an affine $\mathfrak{su}(2)$ algebra (rather than an affine $\mathfrak{u}(2)$ algebra),
since we also divide out the $\mathfrak{u}(1)$ current
given by
\be
U= - \, (N+2)\, \Big(K^{11}+K^{22}+J^{11}+J^{22}\Big)\ .
\ee
Here we have normalised the $\mathfrak{u}(1)$ current so that its spectrum consists of the integers; then we have
\be
U(z)\, U(w) \sim \frac{\kappa}{(z-w)^2}  \ , \qquad \hbox{with} \quad
\kappa = 2 N (N+2) (N+k+2) \ .
\ee
The complete spectrum of coset fields is now generated by the bosonic currents $J^A$ and the fermions
$\psi^{i,\alpha}$ and $\bar\psi^{i,\alpha}$, subject to the condition that they must transform
in the vacuum representation with respect to the currents \eqref{cosetcur}  as well as $U$.

\subsection{The higher spin currents}

Next we want to construct some of the higher spin generators of the ${\cal W}$-algebra explicitly. We begin with the
spin $\frac{3}{2}$-fields that are given as
\be\label{spin32}
G'^{\a\b}=J^{i,\a}\, \bar\psi^{i,\b}\ ,\qquad \bar G'^{\a\b}=\bar{J}^{i,\a}\, \psi^{i,\b} \ ,
\ee
where no normal ordering is required since the bosonic and fermionic fields do not have any non-trivial OPE.
Note that these fields are indeed primary with respect to $\mathfrak{su}(N)_{k+2}$, as follows from
(\ref{suNtrans}), as well as the fact that the fermions also transform in the (anti-)fundamental
representation with respect to this algebra.

The actual supercharges $G^{\alpha\beta}$ of the non-linear $\tilde{A}_{\gamma}$ algebra as well as the
$\tilde{G}^{\pm\pm}$ spin $\frac{3}{2}$-fields that appear in the full ${\cal W}$-algebra are then given by the linear
combinations
\bea
&G^{++}=-2\dfrac{G'^{21}+\bar{G}'^{12}}{\sqrt{k+N+2}}\ ,\qquad\qquad  &\tilde G^{++}=2\dfrac{-G'^{21}+\bar{G}'^{12}}{\sqrt{k+N+2}}\\
&G^{+-}=2\dfrac{-G'^{22}+\bar{G}'^{11}}{\sqrt{k+N+2}}\ ,\qquad \qquad &\tilde G^{+-}=2\dfrac{-G'^{22}-\bar{G}'^{11}}{\sqrt{k+N+2}}\\
& G^{-+}=2\dfrac{G'^{11}-\bar{G}'^{22}}{\sqrt{k+N+2}}\ ,\qquad \qquad & \tilde G^{-+}=2\dfrac{G'^{11}+\bar{G}'^{22}}{\sqrt{k+N+2}}\\
& G^{--}=2\dfrac{G'^{12}+\bar{G}'^{21}}{\sqrt{k+N+2}}\ ,\qquad \qquad & \tilde G^{--}=2\dfrac{G'^{12}-\bar{G}'^{21}}{\sqrt{k+N+2}}\ .
\eea
Both $G^{\alpha\beta}$ and $\tilde{G}^{\alpha\beta}$ transform as primary fields in the $({\bf 2},{\bf 2})$ of the
$\mathfrak{su}(2)_k \oplus \mathfrak{su}(2)_N$ current algebra. We can now calculate their OPEs, and we find for example
\bea
\nn G^{++}(z)\, G^{++}(w)&\sim&-\frac{4}{(k+N+2)(z-w)}\big(((J^+\bar{\psi}^{i1})\psi^{i2})+((J^+{\psi}^{i2})\bar\psi^{i1})\big)\\
&\sim &-\frac{8(J^+K^+)}{(k+N+2)(z-w)}\ ,\label{cftgppgpp}
\eea
where $J^{\pm}$, $J^3$ are the $J$ currents in the Cartan-Weyl basis. (We shall also use a similar notation for the $K$ currents.)
In the 't Hooft limit \eqref{thooft}, eq.~\eqref{cftgppgpp} agrees with the result from the asymptotic symmetry
analysis, eq.~(\ref{GppGpp}).

\noindent Similarly, we find
\bea
\nn G^{++}(z)\, G^{+-}(w)&\sim&\frac{8N J^+}{(k+N+2)(z-w)^2}\nonumber \\
& & \qquad
+\frac{4 \Bigl[ (\bar J^{i1}J^{i2})-(J^{i2}\bar J^{i1})-
\bigl(((J^+\bar{\psi}^{i1})\psi^{i1})+((J^+{\psi}^{i2})\bar\psi^{i2})\big)\bigr)\Bigr]}{(k+N+2)(z-w)} \nonumber\\
\nn&\sim &\frac{8N J^+}{(k+N+2)(z-w)^2}+\frac{4 \Bigl[-N\pa J^{+}-2(-N\pa J^{+})+2(J^{+}K^{3})\Bigr]}{(k+N+2)(z-w)} \nonumber\\
&\sim &\frac{8N J^+}{(k+N+2)(z-w)^2}+\frac{4 \big(N\pa J^{+}+2(J^{+}K^{3})\big)}{(k+N+2)(z-w)}\ , \label{G++G+-}
\eea
where  the second term  $-2(-N\pa J^{+})$ in the second equality comes from the normal ordering
of each of the two terms $\big((J^+\bar{\psi}^{i1})\psi^{i1})+(J^+{\psi}^{i2})\bar\psi^{i2}\big)$, using
the rearrangement identity of normal-ordered-products,
\begin{eqnarray}
\bigl( (AB)C\bigr)-\bigl(A(BC)\bigr) & = & (-1)^{F(B)F(C)}\, \bigl(A([C,B])\bigr)+(-1)^{(F(A)+F(B))F(C)}\,
\bigl(([C,A])B\bigr)\nonumber \\
& & +[(AB),C] \ ,
\end{eqnarray}
where $F(A)=1$ for fermions and $F(A)=0$ for bosons.
In particular, since $[J^{+},\psi^B]=0$ and $[\psi^A,\psi^B]=0$, we have
\be\label{nojpp}
\bigl((J^+\psi^B)\psi^C\bigr)-\bigl(J^+(\psi^B\psi^C)\bigr)=[(J^{+}\psi^B),\psi^C]=\del^{BC}\pa J^{+} \ .
\ee
Again, in the 't Hooft limit, eq.~(\ref{G++G+-}) matches with the result in eq.~\eqref{GppGpm}. The other cases
work similarly, the only complication being the form of the stress-energy tensor in these variables, which takes the somewhat
cumbersome form
\bea
\nn T&=& -\frac{1}{k+N+2}\big((J^{i1}\bar{J}^{i1})+(\bar{J}^{i2}{J}^{i2}) \bigr)+
k(\pa \bar{\psi}^{i,1}\psi^{i1}+ \bar{\psi}^{i,2}\pa\psi^{i2})\\[4pt]
\nn &&-(t^a_\Box)_{ij}\, \bigl((J^a({\psi^{i1}}\bar{\psi}^{j1}))
+(J^a({\psi^{i2}}\bar{\psi}^{j2}))\bigr)\\[4pt]
\nn &&+(t^a_\Box)_{1+N,1+N}\, \bigl((J^a({\psi^{i2}}\bar{\psi}^{i2})\bigr)
+(t^a_\Box)_{2+N,2+N}\, \bigl(J^a({\psi^{i1}}\bar{\psi}^{i1})\big)\\[4pt]
&&+\frac{C^++C^--2(A^{+3}A^{-3})-k \pa A^{-3}+N \pa A^{+3} }{k+N+2} \ , \label{Tope}
\eea
where
\bea
C^+&=&\frac{1}{2}\bigl((A^{++}A^{+-})+(A^{+-}A^{++})\bigr)+(A^{+3}A^{+3})\label{cp}\\
C^-&=&\frac{1}{2}\bigl((A^{-+}A^{--})+(A^{--}A^{-+})\bigr)+(A^{-3}A^{-3})\label{cm}\ .
\eea

\noindent We can similarly study the OPEs of the $\tilde{G}^{\alpha\beta}$ spin-$\frac{3}{2}$ fields by
themselves, and they have exactly the same structure since we have
\begin{equation}
G^{ab}(z)\, G^{cd}(w)=-\tilde G^{ab}(z)\, \tilde G^{cd}(w)\ .
\end{equation}
This mirrors precisely what happens in the asymptotic symmetry algebra.
It is therefore more interesting to analyse the mixed OPEs,
involving both $G^{\alpha\beta}$ and $\tilde{G}^{\alpha\beta}$ fields, in particular
\begin{equation}\label{smlope}
G^{++}(z)\, \tilde{G}^{+-}(w) \ ,\qquad \hbox{and} \qquad G^{++}(z)\, \tilde{G}^{--}(w) \ .
\end{equation}
Based on $\mathfrak{su}(2)\oplus \mathfrak{su}(2)$ symmetry considerations, we expect  the OPE of the
first product to contain composite operators of the form $(V ^ {(1)0}A^{++})$, $\pa A^{++}$ as well
as the field $V^{(2)++}$ of conformal weight $h=2$ in the $R^{(1)}$ multiplet. An explicit computation shows
that $V^{(2)++}$ does indeed appear in the OPE, as there are terms that cannot be written in terms of
$(V ^ {(1)0}A^{++})$ and $\pa A^{++}$ alone.
We can fix the relative coefficients among $(V ^ {(1)0}A^{++})$, $\pa A^{++}$ and $V^{(2)++}$ by requiring
that $V^{(2)++}$ is a Virasoro primary that transforms as the highest weight state of the $({\bf 3},{\bf 1})$ representation of
$\mathfrak{su}(2)\oplus \mathfrak{su}(2)$.
However, this does not fix the normalisation of $V^{(2)++}$ uniquely.

Since we are mainly interested in the comparison with the bulk asymptotic symmetry algebra, we therefore
proceed as follows: we {\em fix} the normalisation of $V^{(2)+\pm}$ by the requirement that the OPE of
$G^{+\pm}(z)\tilde{G}^{+\pm}(w)$ agrees with what was obtained in the asymptotic symmetry algebra analysis
of the higher spin theory, and similarly for $V^{(2)-\pm}$, i.e., we take.
\be
V^{(2)+*}_{\text{CFT}}\sim \frac{iV^{(2)+*}_{\text{bulk}}}{2 a_{(2)++}}\ , \quad \hbox{e.g.} \ \
V^{(2)++}_{\text{CFT}} \equiv \frac{iV^{(2)++}_{\text{bulk}}}{2 a_{(2)++}} \ \hbox{in \eqref{Gppgppm}, \ etc.}
\ee
This then leads to
\bea
\nn V^{(2)++}&=&\frac{4}{k+N+2}\Big((\bar{J}^{i1}J^{i2})+
(J^{i2}\bar{J}^{i1})+\frac{N+2}{k}(J^+(J^{11}+J^{22}))\Big)\\
V^{(2)+-}&=&\frac{4}{k+N+2}\Big((\bar{J}^{i2}J^{i1})+
(J^{i1}\bar{J}^{i2})+\frac{N+2}{k}(J^-(J^{11}+J^{22}))\Big)\\\label{v2p}
\nn V^{(2)+3}&=&\frac{8}{k+N+2}\Big((J^{i2}\bar{J}^{i2})-(J^{i1}\bar{J}^{i1})
-\frac{N+2}{k}(J^3(J^{11}+J^{22}))+N \pa J^3\Big) \ .
\eea
(Obviously, we should mention that the normalisation of these expressions is
only determined by this procedure up to subleading terms
in the 't~Hooft limit.) Similarly, by the same method we also fix the normalisation of the spin-$1$ current $V^{(1) 0}$ that
appears as
the leading singularity in the OPE of  $G^{++}G'^{--}$
\be
V^{(1) 0}=\frac{1}{(k+N+2)}\Big(k\, (K^{11}+K^{22})-(N+2)\, (J^{11}+J^{22})\Big)\ .
\ee
Now that we have fixed these various definitions, we can unambiguously calculate the OPEs
\be\label{gppVppc}
G^{++}(z)\, V^{(2)++}(w)\sim \frac{4\frac{N+2k+2}{k(k+N+2)}(A^{++}G'^{++})}{z-w}\ ,
\ee
as well as
\be\label{VppVppc}
V^{(2)++}(z)\, V^{(2)++}(w)=\frac{\frac{-32 N\left(2 k+N+2 \right)}{k (k+N+2)^2}(A^{++}
A^{++})(w)}{(z-w)^2}
-\frac{\frac{32 N (2 k+N+2)}{k (k+N+2)^2}(A^{++}\pa A^{++})(w)}{(z-w)}\ .
\ee
It is now a real consistency check that they agree with eqs.\
\eqref{gppVpp} and \eqref{VppVpp}, respectively, of the asymptotic symmetry algebra in the
't~Hooft limit.

\noindent Similarly, we can compare the leading terms of the OPE
\bea
\nn & V^{(2)++}(z) & \hspace*{-0.2cm}  V^{(2)+-}(w)\sim -\tfrac{32 (k-1) N (2 k+N+2)}{(k+N+2)^2} \frac{1}{(z-w)^4}
-\tfrac{(64 (-1 + k) N (2 + 2 k + N)  }{k (2 + k + N)^2 }\, \frac{A^{+3}}{(z-w)^3}  \\[2pt]
\nn && +\tfrac{16 (-1 + k) (1 + k) (2 + N) (2 + 2 k + N)}{k (-(2 + N)^2 +
   k^2 (-1 + 4 N + 3 N^2) + k (-4 + 3 N + 4 N^2))}\frac{( V^{(1) 0} \, V^{(1) 0} \,)}{(z-w)^2}\\[2pt]
\nn &&
-\tfrac{32 (-1 + k) N (2 + 2 k + N)}{k (2 + k + N)^2}\frac{\pa A^{+3}}{(z-w)^2}
-\tfrac{32 N (2 + 2 k + N)}{k (2 + k + N)^2}\frac{(A^{+3}A^{+3})}{(z-w)^2}\\[2pt]
\nn  &&+\tfrac{64 (-1 + k) (1 + k) N (2 + 2 k + N)}{(2 + k + N) (-(2 + N)^2 +
   k^2 (-1 + 4 N + 3 N^2) + k (-4 + 3 N + 4 N^2))}\frac{C^-}{(z-w)^2}\\[2pt]
\nn   && -\tfrac{32 N (2 + 2 k + N) (-2 k^3 (2 + N) + (2 + N)^2 +
    k (4 - 3 N - 4 N^2) + k^2 (-1 + N^2)}{
 k (2 + k + N)^2 (-(2 + N)^2 + k^2 (-1 + 4 N + 3 N^2) +
    k (-4 + 3 N + 4 N^2))}\frac{C^+}{(z-w)^2}\\[2pt]
\nn &&-\tfrac{64 (-1 + k) (1 + k) N (2 + N) (2 + 2 k + N)}{(2 + k +
    N) (-(2 + N)^2 + k^2 (-1 + 4 N + 3 N^2) +
    k (-4 + 3 N + 4 N^2))}\frac{ T-V^{(2)0}}{(z-w)^2}\\[2pt]
&&+\co\bigl((z-w)^{-1} \bigr)\ ,\label{vss}
\eea
where $C^+$ and $C^-$ are defined in eqs.~\eqref{cp} and \eqref{cm}, respectively, and we have fixed
the normalisation of $V^{(2)0}$ (up to a sign) by the requirement that
\be
V^{(2)0}(z)\, V^{(2)0}(w)\sim \frac{\frac{k (N-1) (N+1) (k+2 N+2)
\left(3 k^2 N^2+4 k^2 N-k^2+4 k N^2+3 k N-4 k-N^2-4 N-4\right)}{(k-1) (k+1) N (N+2)^2 (k+N+2) (2 k+N+2)}}{(z-w)^4}+\cdots \  ,
\ee
where the ellipses stand for subleading terms, see (\ref{v2kv2k}). Again, it  is now a non-trivial consistency check that all the OPE
coefficients in (\ref{vss}) agree, in the 't~Hooft limit \eqref{thooft}, with the corresponding terms in \eqref{VppVpm} from the
asymptotic symmetry algebra.

We can also construct the fields in the $({\bf 1},{\bf 3})$ representation, following the same idea as in \eqref{v2p}
\begin{eqnarray*}
V^{(2)-+}&=&k(\psi^{i,2}\pa \bar{\psi}^{i,1}-\pa \psi^{i,2}\bar{\psi}^{i,1})-(J^{11}(\psi^{i,2}\bar{\psi}^{i,1}))
-(J^{22}(\psi^{i,2}\bar{\psi}^{i,1})) \\
& & +2 \bigl(J^A(\psi^{i,2}\bar{\psi^{j,1}})\bigr)\, (t^A_\Box)_{ij} -\tfrac{2+k+N}{N}\, (V^{(1)0} K^+) \ , \\[4pt]
V^{(2)--}&=&k(\psi^{i,1}\pa \bar{\psi}^{i,2}-\pa \psi^{i,1}\bar{\psi}^{i,2})-(J^{11}(\psi^{i,1}\bar{\psi}^{i,2}))
-(J^{22}(\psi^{i,1}\bar{\psi}^{i,2}))\\
& & +2 \bigl(J^A(\psi^{i,1}\bar{\psi^{j,2}})\bigr)\, (t^A_\Box)_{ij} -\tfrac{2+k+N}{N}\, (V^{(1)0} K^-)\ , \\[4pt]
V^{(2)-3}&=&\tfrac{k(N-1)}{N}\, \bigl(\psi^{i,1}\pa \bar{\psi}^{i,1}-\pa \psi^{i,1}\bar{\psi}^{i,1}
-\psi^{i,2}\pa \bar{\psi}^{i,2}+\pa \psi^{i,2}\bar{\psi}^{i,2}\bigr)\\
&&+2(J^{11}(\psi^{i,1}\bar{\psi}^{i,1}))
+2(J^{22}(\psi^{i,1}\bar{\psi}^{i,1}))
-2(J^{11}(\psi^{i,2}\bar{\psi}^{i,2}))
-2(J^{22}(\psi^{i,2}\bar{\psi}^{i,2}))\\
&&+2 \bigl(J^A(\psi^{i,1}\bar{\psi^{j,1}})\bigr)\, (t^A_\Box)_{ij}-2 \bigl(J^A(\psi^{i,2}\bar{\psi^{j,2}})\bigr)\, (t^A_\Box)_{ij}\\
&&+\tfrac{k}{N} \bigl((\psi^{i2}\psi^{i2})(\psi^{2}\psi^{j2})-(\psi^{i1}\psi^{i1})(\psi^{j1}\psi^{j1})\bigr) \ .
\end{eqnarray*}
Some of the OPEs involving these operators are then
\be
G^{++}(z)\, V^{(2)-+}(w)\sim \frac{-4\frac{2N+k+2}{N(k+N+2)}(A^{-+}G'^{++})}{z-w}\ ,\label{GppVmp}
\ee
as well as
\be
V^{(2)-+}(z)\, V^{(2)-+}(w)=\frac{-\frac{32 k (k+2 N+2)}{N (k+N+2)^2}(A^{-+}
A^{-+})(w)}{(z-w)^2}
-\frac{\frac{32 k (k+2 N+2)}{N (k+N+2)^2}(A^{-+}\pa A^{-+})(w)}{(z-w)}\ .\label{VmpVmp}
\ee
It is again straightforward to check that they match the coefficients of the asymptotic symmetry algebra (see the ancillary file) in
the 't~Hooft limit.

\subsection{Truncation of the ${\cal W}_\infty$ algebra}\label{wtruncation}

As we have seen in section~\ref{astruncation}, the (classical) asymptotic symmetry algebra associated to
${\rm shs}_2[\m]$ truncates for $\mu=s+1$ or $\mu=-s$. It is then natural to suspect that
the same truncation will also happen in the actual quantum ${\cal W}_{\infty}$ algebra. (Since the
asymptotic symmetry algebra is non-linear, the transition from the classical Poisson algebra to the
quantum Lie algebra is non-trivial, see \cite{Gaberdiel:2012ku}.) While we have not yet studied
the most general ${\cal W}_{\infty}$ algebra with this spin content ---  this will be done in
\cite{BCG} --- we can at least analyse this question for the above cosets.

First of all, we can check whether the relevant multiplet shortens at the appropriate value of $(k^+,k^-)$, i.e.,
whether a certain descendant of the $\tilde{A}_\gamma$ primary spin $s$ field $\Phi_s$ becomes null.
One finds that\footnote{This point
was missed in \cite{Gaberdiel:2013vva}, where it was only observed
that the first term does {\em not} become null, see eq.~(2.43) of that paper. Note that the
coset algebra also makes sense for non-positive level $k^+$, so these conditions can be satisfied.}
\begin{eqnarray}
\cn^+= \Bigl( G^{+-}_{-\frac{1}{2}}G^{++}_{-\frac{1}{2}}+\frac{4s}{k^+}A^{++}_{-1} \Bigr)\, \F_s  \cong 0 \qquad
\hbox{for $sk^-+(1+s)k^++2s=0\ $ or $\ k^+=1$}\ ,  \label{nullm} \\
\cn^-=\Bigl( G^{++}_{-\frac{1}{2}}G^{-+}_{-\frac{1}{2}} -\frac{4s}{k^-}A^{-+}_{-1}\Bigr)\, \F_s\cong 0 \qquad
\hbox{for $sk^++(1+s)k^-+2s=0\ $  or $\ k^-=1$} \ ,
\label{nullp}
\end{eqnarray}
i.e., that these states are annihilated by the positive $\tilde{A}_\gamma$ generators.
The first solution in each of these cases corresponds to
\begin{eqnarray}
sk^-+(1+s)k^++2s=0 \qquad & \Longrightarrow & \quad  \gamma \equiv \frac{k^-}{k^+ +k^-} = \frac{ (s+1) k^-}{k^- - 2 s}\,, \\[2pt]
sk^++(1+s)k^-+2s=0 \qquad & \Longrightarrow & \quad \gamma \equiv \frac{k^-}{k^+ +k^-} = - \frac{s k^-}{k^-+2s} \ ,
\end{eqnarray}
i.e., in the first case we get $\gamma = s+1$ in the 't~Hooft limit, while in the second case the asymptotic value is
$\gamma = -s$, in nice agreement with what we saw in section~\ref{astruncation}. (Note that
${\cal N}^+$ and ${\cal N}^-$ transform in the $({\bf 3},{\bf 1})$ and $({\bf 1},{\bf 3})$ representations, respectively.)
Incidentally, we should note that the two cases correspond precisely to
\begin{eqnarray}
& sk^-+(1+s)k^++2s=0 \ :  \qquad  & \gamma_2 = \frac{k^+}{k^+ + k^- + 2} = -s  \\[2pt]
& sk^++(1+s)k^-+2s=0 \ :  \qquad  & \gamma_1 =  \frac{k^-}{k^+ + k^- + 2} = -s  \ ,
\end{eqnarray}
where $\gamma_1$ and $\gamma_2$ were introduced in eq.~(B.38) of \cite{Gaberdiel:2013vva}.

We can also check whether some of the other higher spin fields decouple appropriately for these values of
$(k^+,k^-)$. As in section~\ref{astruncation}, let us consider explicitly the example $s=1$ where we expect that
for $\g_2=-1$, i.e., for $k^-+2k^++2=0$, the generators $ V^{(2)+*}$ decouple. In the above coset description
we have $k^+=k$ and $k^-=N$, and thus for example the right-hand-side of \eqref{gppVppc} and \eqref{VppVppc}
should vanish for $N+2k+2=0$ --- which it indeed does. Similarly, for $\g_1=-1$ or $k^++2k^-+2=0$, the generators
$V^{(2)-*}$ should belong to the ideal (while the generators $ V^{(2)+*}$  survive). Looking at the structure of
the OPEs  \eqref{GppVmp} and \eqref{VmpVmp}, this is indeed the case. These two examples provide
non-trivial evidence for the assertion that the quantum ${\cal W}_\infty$ algebra indeed truncates in the same
way as the (classical) asymptotic symmetry algebra.
\smallskip

\noindent For the second solution in (\ref{nullm}) and (\ref{nullp}), we find
\begin{eqnarray}
k^+=1 \qquad & \Longrightarrow & \quad  \gamma \equiv \frac{k^-}{k^+ +k^-} = \frac{k^-}{k^- +1}\ , \\[2pt]\label{simpnp}
k^-=1 \qquad & \Longrightarrow & \quad  \gamma \equiv \frac{k^-}{k^+ +k^-} = \frac{1}{k^+ +1}\ , \label{simpnm}
\end{eqnarray}
which corresponds, in the 't~Hooft limit, to $\gamma= 1$ and $\gamma=0$, respectively. In this limit, one of the
two $\mathfrak{su}(2)$ current algebras decouples, and the large $\cn=4$ superconformal algebra becomes the small $\cn=4$ superconformal algebra.
Finally, we note that for
\be\label{toosp}
k^+ = k^- = -\frac {2 s} {1 + 2 s}\quad \text{ or } \quad k^+ = k^- = 1
\ee
both vectors $\cn^+$ and $\cn^-$ become simultaneously null. The solution $k^+=k^-$ corresponds to
$\gamma = \frac{1}{2}$, where
$D(2,1|\a)$ is isomorphic to ${\rm OSp}(4|2)$; this truncation may therefore play a role in connecting
the large $\cn=4$ $\cw_\infty$ algebra to the ${\rm shs}^E(N|2)$ algebra based on ${\rm OSp}(N|2)$
that was studied in \cite{Henneaux:2012ny}.
\smallskip

A similar analysis can also be done for the linear $A_\gamma$ algebra, see appendix~\ref{agammanull} for details.
The non-linear $\tilde{A}_\g$ algebra can be obtained from $A_\g$ upon factoring out the free fermions and the
$\mathfrak{u}(1)$ generator, and this process reduces the levels of the $\mathfrak{su}(2)$ algebras by one,
$k^\pm \mapsto k^\pm -1$. One may therefore expect that the conditions for the appearance of the null-vectors are
related to the above by this shift, and this is indeed borne out by the analysis of appendix~\ref{agammanull}.

\section{The perturbation analysis}\label{sec:perturbation}

In this section we use the results from the previous section to study the effect of the perturbation by the
$({\rm f};\bar{\rm f})$ field on the higher spin currents.
Recall from \cite{Gaberdiel:2013vva} (see the discussion around eq.~(4.12) of that paper)
that the ground state of the coset representation $({\rm f};\bar{\rm f})$ is to be identified in the above
notation with the state
\be\label{ffbar}
\bar{\psi}^{\,i,\alpha}_{-1/2} \, |{(\bf N+2}) \rightarrow {\bf 2} \rangle \ ,
\ee
where $|({\bf N+2}) \rightarrow {\bf 2} \rangle$ denotes the states in the fundamental representation
of $\mathfrak{su}(N+2)$ that transform with respect to the branching of
$\mathfrak{su}(N+2)$ into $\mathfrak{su}(N) \oplus \mathfrak{su}(2) \oplus \mathfrak{u}(1)$
\be\label{fundb}
({\bf N+2}) = {\bf N}_2 \oplus {\bf 2}_{-N}
\ee
as a singlet of $\mathfrak{su}(N)$ and a ${\bf 2}$ of $\mathfrak{su}(2)$. (Here the indices in (\ref{fundb})
denote the $\mathfrak{u}(1)$ charge; note that our $\mathfrak{u}(1)$ generator has been rescaled by
a factor of $2$ relative to \cite{Gaberdiel:2013vva} so that all its eigenvalues are integers.) The state
(\ref{ffbar}) has conformal dimenion $h=\tfrac{1}{2}$, and it transforms in the doublet representation
of both $\mathfrak{su}(2)$ algebras of $\tilde{A}_\gamma$. It therefore defines a chiral primary operator,
and its $G^{**}_{-1/2}$ descendant gives rise to an exactly marginal pertubation that preserves
the large ${\cal N}=4$ superconformal algebra. The actual perturbing state then has the form
\be
\Phi = \bar{J}^{\,i,\alpha}_{-1} \, |{(\bf N+2}) \rightarrow {\bf 2} \rangle \ ,
\ee
as follows from the structure of the supercharges (\ref{spin32}).
\smallskip

The question of whether the higher spin currents are preserved by the perturbation now boils
down to an analysis of the poles in the OPE of $\Phi$ with the higher spin currents, see
section~4 of \cite{Gaberdiel:2013jpa} for a discussion in a related context. The simplest
case to study is the spin $1$ current $V^{(1) 0}$ that sits at the bottom of the $R^{(1)}$ multiplet.
In this case, the analysis of  \cite{Gaberdiel:2013jpa} implies that
$V^{(1) 0}$ is preserved by the perturbation provided that the
eigenvalue of $V^{(1) 0}_0$ on $\Phi$ vanishes.
Given that
$\Phi$ does not involve any fermions, the eigenvalue under $K^{11}+K^{22}$ is
obviously trivial; on the other hand, from the fact that the eigenvalue of $\Phi$ under $U_0$
equals $- 2 (N+1)$, we conclude that\footnote{This can also be confirmed independently.}
\be
V^{(1) 0}_0 \, \Phi =  - \frac{2  (N+1)}{(k+N+2)} \, \Phi \neq 0 \ .
\ee
Thus it follows that the spin $1$ current $V^{(1) 0}$ is not preserved by the perturbation. We should
mention that while we have performed this analysis for the version of the coset theory that
contains the non-linear $\tilde{A}_\gamma$ algebra, the result is identical also in the other case
(that contains the linear $A_\gamma$ algebra) since the spin $1$ generator $V^{(1) 0}$ is unmodified
in going from one description to the other.

Since the OPEs of $R^{(1)}$ generate recursively the full ${\cal W}_{\infty}$ algebra, this result then
suggests that in fact all higher spin currents (outside the large ${\cal N}=4$ superconformal
algebra) will acquire anomalous dimensions under this perturbation, and hence that only the
superconformal symmetry survives.

\section{Conclusion}

In this paper we have checked that the asymptotic symmetries of the AdS$_3$ higher spin theory
based on ${\rm shs}_2[\lambda]$ match those of the 2d  CFT Wolf space cosets in the
't~Hooft limit. This provides an important and non-trivial confirmation for the ${\cal N}=4$
duality that was proposed in \cite{Gaberdiel:2013vva}. We have performed our analysis for the `non-linear'
version of the duality where the large ${\cal N}=4$ superconformal algebra is a non-linear
${\cal W}$-algebra on both sides. As we have also explained, by introducing auxiliary boundary degrees
of freedom, we can partially linearise the asymptotic symmetry algebra, so that it contains
the linear large ${\cal N}=4$ superconformal algebra as a subalgebra (thereby matching a
similar construction on the coset side). The two versions of the duality (linear vs.\ non-linear)
differ only by the inclusion of finitely many free boundary fields; their difference is therefore invisible
in the classical large central charge limit, and hence corresponds to an ambiguity in
characterising the (semiclassical) asymptotic symmetry algebra of the higher spin theory.

In the process of checking the agreement of the symmetries, we have also worked out explicit
expressions for the first few ${\cal W}$-algebra generators for the (non-linear) Wolf coset CFTs.
In turn, these expressions allowed us to determine the behaviour of the ${\cal W}_\infty$
symmetry under exactly marginal perturbations. In particular, we have shown that the
spin $1$ generator of the first non-trivial ${\cal N}=4$ multiplet becomes anomalous under
the exactly marginal ${\cal N}=4$ preserving deformation of \cite{Gaberdiel:2013vva}. This suggests that the entire
higher symmetry will get broken by this perturbation. This should help to shed light on the
interpretation of this perturbation from the dual higher spin theory viewpoint.

\appendix

\section*{Acknowledgements}

We thank Constantin Candu, Rajesh Gopakumar and Carl Vollenweider for useful discussions. This
research is supported in parts by a grant from the Swiss National Science Foundation. MRG thanks
the Weizmann Institute in Rehovot for hospitality during the final stages of this work.

\newpage

\section{Details of the asymptotic symmetry algebra}\label{opeapd}
In the following we provide some OPEs of the asymptotic symmetry algebra;
additional OPEs can be found in the ancillary file on the {\tt arXiv}.
In various places, we have introduced some normalisation parameters (that are usually denoted by
$a_{(s)**}$). We can simply set them to arbitrary values representing different normalisation conventions for the generators.

{\footnotesize
\begin{eqnarray}
G^{++}(z)G'^{++}(w) &\sim&  0 \nn\\[2pt]
G^{++}(z)G'^{+-}(w) &\sim& \frac{4 \gamma  A^{++}(w)\,V ^ {(1)0}(w)}{k_{\rm cs} (w-z)}+\frac{i V^{(2)++}(w)}{2 a_{(2)++} (w-z)} \nn \\[2pt]
G^{++}(z)G'^{-+}(w) &\sim& -  \frac{4 (\gamma -1) A^{-+}(w)\,V ^ {(1)0}(w)}{k_{\rm cs} (w-z) }-\frac{i V^{(2)-+}(w)}{2 a_{(2)-+} (w-z)} \nn  \\[2pt]
G^{++}(z)G'^{--}(w) &\sim& \frac{i \big(8 i a_{(2)-+} a_{(2)++} \pa V ^ {(1)0}(w)-\sqrt{2} a_{(2)++} V^{(2)-3}(w)
+\sqrt{2} a_{(2)-+} V^{(2)+3}(w)\big)}{4 (w-z) a_{(2)-+} a_{(2)++}} \nn \\
&&+\frac{2 (2 \gamma -2) A^{-3}(w)\,V ^ {(1)0}(w)}{k_{\rm cs} (w-z)}-\frac{4 \gamma  A^{+3}(w)\,V ^ {(1)0}(w)}{k_{\rm cs} (w-z)}+\frac{4 V ^ {(1)0}(w)}{(w-z)^2} \nn \\[2pt]
  G^{+-}(z)G'^{++}(w) &\sim& -\frac{4 \gamma  A^{++}(w)\,V ^ {(1)0}(w)}{k_{\rm cs} (w-z)}-\frac{i V^{(2)++}(w)}{2 a_{(2)++} (w-z)} \nn\\[2pt]
  G^{+-}(z)G'^{+-}(w) &\sim&  0\nn\\[2pt]
  G^{+-}(z)G'^{-+}(w) &\sim&  \frac{8 a_{(2)-+} a_{(2)++} \pa V ^ {(1)0}(w)+\sqrt{2} (-i) a_{(2)++} V^{(2)-3}(w)
  -i \sqrt{2} a_{(2)-+} V^{(2)+3}(w)}{4 (w-z) a_{(2)-+} a_{(2)++}} \nn \\
  &&+\frac{4 (\gamma -1) A^{-3}(w)\,V ^ {(1)0}(w)}{k_{\rm cs} (w-z)}+\frac{4 \gamma  A^{+3}(w)\,V ^ {(1)0}(w)}{k_{\rm cs} (w-z)}-\frac{4 V ^ {(1)0}(w)}{(w-z)^2} \nn\\[2pt]
  G^{+-}(z)G'^{--}(w) &\sim& \frac{2 (2 \gamma -2) A^{--}(w)\,V ^ {(1)0}(w)}{k_{\rm cs}(w-z)}-\frac{i V^{(2)--}(w)}{2 a_{(2)-+}(w-z)} \nn \\[2pt]
 \nn G^{-+}(z)G'^{++}(w) &\sim& \frac{4 (\gamma -1) A^{-+}(w)\,V ^ {(1)0}(w)}{k_{\rm cs}(w-z)}+\frac{i V^{(2)-+}(w)}{2 a_{(2)-+}(w-z)} \nn \\[2pt]
\nn  G^{-+}(z)G'^{+-}(w) &\sim& \frac{8 a_{(2)-+} a_{(2)++} \pa V ^ {(1)0}(w)+\sqrt{2} i a_{(2)++} V^{(2)-3}(w)
+\sqrt{2} i a_{(2)-+} V^{(2)+3}(w)}{4 (w-z) a_{(2)-+} a_{(2)++}}\\
  &&-\frac{4 (\gamma -1) A^{-3}(w)\,V ^ {(1)0}(w)}{k_{\rm cs} (w-z)}-\frac{4 \gamma  A^{+3}(w)\,V ^ {(1)0}(w)}{k_{\rm cs} (w-z)}-\frac{4 V ^ {(1)0}(w)}{(w-z)^2}
  \label{Gppgppm}\\[2pt]
\nn  G^{-+}(z)G'^{-+}(w) &\sim&  0\\[2pt]
  G^{-+}(z)G'^{--}(w) &\sim&  -\frac{4 \gamma  A^{+-}(w)\,V ^ {(1)0}(w)}{k_{\rm cs}(w-z)}+\frac{i V^{(2)+-}(w)}{2 a_{(2)++}(w-z)} \\[2pt]
\nn  G^{--}(z)G'^{++}(w) &\sim&  -\frac{i \big(-8 i a_{(2)-+} a_{(2)++} \pa V ^ {(1)0}(w)-\sqrt{2} a_{(2)++} V^{(2)-3}(w)
+\sqrt{2} a_{(2)-+} V^{(2)+3}(w)\big)}{4 (w-z) a_{(2)-+} a_{(2)++}}\\
\nn  &&-\frac{4 (\gamma -1) A^{-3}(w)\,V ^ {(1)0}(w)}{k_{\rm cs} (w-z)}+\frac{4 \gamma  A^{+3}(w)\,V ^ {(1)0}(w)}{k_{\rm cs} (w-z)}+\frac{4 V ^ {(1)0}(w)}{(w-z)^2}\\[2pt]
\nn  G^{--}(z)G'^{+-}(w) &\sim& -\frac{4 (\gamma -1) A^{--}(w)\,V ^ {(1)0}(w)}{k_{\rm cs}(w-z)}+\frac{i V^{(2)--}(w)}{2 a_{(2)-+}(w-z)} \\[2pt]
\nn  G^{--}(z)G'^{-+}(w) &\sim&  \frac{4 \gamma  A^{+-}(w)\,V ^ {(1)0}(w)}{k_{\rm cs}(w-z)}-\frac{i V^{(2)+-}(w)}{2 a_{(2)++}(w-z)} \\[2pt]
G^{--}(z)G'^{--}(w) &\sim&  0
\end{eqnarray}}
{\footnotesize
\begin{eqnarray}
G'^{++}(z)G'^{++}(w) &\sim& \frac{8 (\gamma -1) \gamma  A^{-+}(w)\,A^{++}(w)}{k_{\rm cs} (w-z)} \nn \\
  G'^{++}(z)G'^{+-}(w) &\sim& -\frac{8 (\gamma -1) \gamma  A^{-3}(w)\,A^{++}(w)}{k_{\rm cs} (w-z)}
  +\frac{4 \gamma  \pa A^{++}(w)}{w-z}-\frac{8 \gamma  A^{++}(w)}{(w-z)^2} \nn \\
  G'^{++}(z)G'^{-+}(w) &\sim& -\frac{8 (\gamma -1) \gamma  A^{-+}(w)\,A^{+3}(w)}{k_{\rm cs} (w-z)}
  -\frac{4 (\gamma -1) \pa A^{-+}(w)}{w-z}+\frac{8 (\gamma -1) A^{-+}(w)}{(w-z)^2} \nn \\[2pt]
  G'^{++}(z)G'^{--}(w) &\sim& \frac{8 (\gamma -1) \gamma  A^{-3}(w)\,A^{+3}(w)}{k_{\rm cs} (w-z)}
  -\frac{2 (\gamma -1) \gamma  A^{--}(w)\,A^{-+}(w)}{k_{\rm cs} (w-z)} \nn \\
& &  -\frac{2 (\gamma -1) \gamma  A^{-+}(w)\,A^{--}(w)}{k_{\rm cs} (w-z)}
-\frac{4 (\gamma -1) \gamma  A^{-3}(w)\,A^{-3}(w)}{k_{\rm cs} (w-z} \nn \\
& & -\frac{2 (\gamma -1) \gamma  A^{+-}(w)\,A^{++}(w)}{k_{\rm cs} (w-z)}
-\frac{2 (\gamma -1) \gamma  A^{++}(w)\,A^{+-}(w)}{k_{\rm cs} (w-z)} \nn \\
&& -\frac{4 (\gamma -1) \gamma  A^{+3}(w)\,A^{+3}(w)}{k_{\rm cs} (w-z)} \nn \\
& & -\frac{4 \big(-\gamma  \pa A^{-3}(w)+\gamma  \pa A^{+3}(w)+\pa A^{-3}(w)+T(w)\big)}{w-z} \nn \\
&&-\frac{8 (\gamma  A^{-3}(w)-\gamma  A^{+3}(w)-A^{-3}(w))}{(w-z)^2}-\frac{8 k_{\rm cs} }{(w-z)^3} \nn \\[2pt]
G'^{+-}(z)G'^{+-}(w) &\sim&  -\frac{8 (\gamma -1) \gamma  A^{--}(w)\,A^{++}(w)}{k_{\rm cs} (w-z)} \nn \\
  G'^{+-}(z)G'^{-+}(w) &\sim&  \frac{2 (\gamma -1) \gamma  A^{--}(w)\,A^{-+}(w)}{k_{\rm cs} (w-z)}
  +\frac{2 (\gamma -1) \gamma  A^{-+}(w)\,A^{--}(w)}{k_{\rm cs} (w-z)} \nn \\
  & & \frac{4 (\gamma -1) \gamma  A^{-3}(w)\,A^{-3}(w)}{k_{\rm cs} (w-z)}
  +\frac{8 (\gamma -1) \gamma  A^{-3}(w)\,A^{+3}(w)}{k_{\rm cs} (w-z)}\nn \\
  & & +\frac{2 (\gamma -1) \gamma  A^{+-}(w)\,A^{++}(w)}{k_{\rm cs} (w-z)}
  +\frac{2 (\gamma -1) \gamma  A^{++}(w)\,A^{+-}(w)}{k_{\rm cs} (w-z)} \nn \\
  &&+\frac{4 (\gamma -1) \gamma  A^{+3}(w)\,A^{+3}(w)}{k_{\rm cs} (w-z)} \nn \\
  &&   +\frac{4 \big(\gamma  \pa A^{-3}(w)+\gamma  \pa A^{+3}(w)-\pa A^{-3}(w)+T(w)\big)}{w-z} \nn \\
  &&-\frac{8 (\gamma  A^{-3}(w)+\gamma  A^{+3}(w)-A^{-3}(w))}{(w-z)^2}+\frac{8 k_{\rm cs} }{(w-z)^3} \nn \\[4pt]
  G'^{+-}(z)G'^{--}(w) &\sim& \frac{8 (\gamma -1) \gamma  A^{--}(w)\,A^{+3}(w)}{k_{\rm cs} (w-z)}
  +\frac{4 (\gamma -1) \pa A^{--}(w)}{w-z}-\frac{8 (\gamma -1) A^{--}(w)}{(w-z)^2} \nn \\
  G'^{-+}(z)G'^{-+}(w) &\sim& -\frac{8 (\gamma -1) \gamma  A^{-+}(w)\,A^{+-}(w)}{k_{\rm cs} (w-z)} \nn \\
  G'^{-+}(z)G'^{--}(w) &\sim&  \frac{8 (\gamma -1) \gamma  A^{-3}(w)\,A^{+-}(w)}{k_{\rm cs} (w-z)}
  -\frac{4 \gamma  \pa A^{+-}(w)}{w-z}+\frac{8 \gamma  A^{+-}(w)}{(w-z)^2}\nn \\
 G'^{--}(z)G'^{--}(w) &\sim&  \frac{8 (\gamma -1) \gamma  A^{--}(w)\,A^{+-}(w)}{k_{\rm cs} (w-z)}\nn \nn \\
 G^{++}(z)V^{(2) 0}(w) &\sim&  \frac{3 (-1)^{3/4} \big(G^{(5/2)-1}(w)+i G^{(5/2)-2}(w)+G^{(5/2)+1}(w)+i G^{(5/2)+2}(w)\big)}{16 (\gamma -2) (w-z)} \nn \\
  G^{+-}(z)V^{(2) 0}(w) &\sim&  \frac{3 (-1)^{3/4} \big(G^{(5/2)-0}(w)-G^{(5/2)-3}(w)-G^{(5/2)+0}(w)-G^{(5/2)+3}(w)\big)}{16 (\gamma -2) (w-z)}\nn \\
  G^{-+}(z)V^{(2) 0}(w) &\sim& -\frac{3 (-1)^{3/4} \big(G^{(5/2)-0}(w)+G^{(5/2)-3}(w)-G^{(5/2)+0}(w)+G^{(5/2)+3}(w)\big)}{16 (\gamma -2) (w-z)} \nn \\
  G^{--}(z)V^{(2) 0}(w) &\sim& -\frac{3 \sqrt[4]{-1} \big(i G^{(5/2)-1}(w)+G^{(5/2)-2}(w)+i G^{(5/2)+1}(w)+G^{(5/2)+2}(w)\big)}{16 (\gamma -2) (w-z)}\nn \ .
\end{eqnarray}}

\noindent Here $G^{(5/2)\pm j}$ are a set of spin-$\frac{5}{2}$ fields, but we have not attempted to write them in a particular
basis. For the comparison to the CFT calculation we also need the OPEs
\be
G^{++}(z)\, V^{(2)++}(w) \sim  -\frac{8 i (\gamma -2) \gamma  a_{(2)++} A^{++}(w)\,G^{'++}(w)}{k_{\rm cs} (w-z)}\label{gppVpp}
\ee
\bea
V^{(2)++}(z)\, V^{(2)++}(w) & \sim  & \frac{128 (\gamma -2) \gamma ^2 a_{(2)++}^2 A^{++}(w)\,\pa A^{++}(w)}{k_{\rm cs} (w-z)} \nn \\
& & -\frac{128 (\gamma -2) \gamma ^2 a_{(2)++}^2 A^{++}(w)\,A^{++}(w)}{k_{\rm cs} (w-z)^2}\label{VppVpp} \ ,
\eea
as well as
{\footnotesize
\begin{eqnarray}
\nn &&\hspace{-10mm}V^{(2)++}(z)V^{(2)+-}(w) \\
&\sim& \frac{128 (\gamma -2) (\gamma -1) A^{--}(w)\,\pa A^{-+}(w) a_{(2)++}^2}{3 k_{\rm cs} (w-z)}
+\frac{128 (\gamma -2) (\gamma -1) A^{-+}(w)\,\pa A^{--}(w) a_{(2)++}^2}{3 k_{\rm cs} (w-z)} \nn \\
\nn  &&+\frac{256 (\gamma -2) (\gamma -1) A^{-3}(w)\,\pa A^{-3}(w) a_{(2)++}^2}{3 k_{\rm cs} (w-z)}
+\frac{128 (\gamma -2) \gamma  (2 \gamma -1) A^{+-}(w)\,\pa A^{++}(w) a_{(2)++}^2}{3 k_{\rm cs} (w-z)}\\
\nn  &&+\frac{128 (\gamma -2) (\gamma -1) \gamma  A^{++}(w)\,\pa A^{+-}(w) a_{(2)++}^2}{3 k_{\rm cs} (w-z)}+\frac{512 (\gamma -2) \gamma  A^{+3}(w)\,T(w) a_{(2)++}^2}{3 k_{\rm cs} (w-z)}\\
\nn  &&+\frac{256 (\gamma -2) \gamma  (3 \gamma -1) A^{+3}(w)\,\pa A^{+3}(w) a_{(2)++}^2}{3 k_{\rm cs} (w-z)}+\frac{64 (\gamma -2) V ^ {(1)0}(w)\,V ^ {(1)0}(w) a_{(2)++}^2}{3 k_{\rm cs} (w-z)^2}\\
\nn  &&+\frac{16 (\gamma -2) (5 \gamma -7) G^{-+}(w)\,G^{+-}(w) a_{(2)++}^2}{9 k_{\rm cs} (w-z)}+\frac{64 (\gamma -2)^2 G^{'--}(w)\,G^{'++}(w) a_{(2)++}^2}{9 k_{\rm cs} (w-z)}\\
\nn  &&+\frac{512 (\gamma -2) (\gamma -1) \gamma  A^{--}(w)\,A^{-+}(w)\,A^{+3}(w) a_{(2)++}^2}{3 k_{\rm cs}^2 (w-z)} \\
& & \nn +\frac{512 (\gamma -2) (\gamma -1) \gamma  (A^{-3}(w))^2\,A^{+3}(w) a_{(2)++}^2}{3 k_{\rm cs}^2 (w-z)}\\
\nn  &&+\frac{512 (\gamma -2) (\gamma -1) \gamma ^2 A^{+-}(w)\,A^{++}(w)\,A^{+3}(w) a_{(2)++}^2}{3 k_{\rm cs}^2 (w-z)}
+\frac{512 (\gamma -2) (\gamma -1) \gamma ^2 (A^{+3}(w))^2\,A^{+3}(w) a_{(2)++}^2}{3 k_{\rm cs}^2 (w-z)}\\
\nn  &&+\frac{256 (\gamma -2) \gamma  A^{+3}(w) a_{(2)++}^2}{(w-z)^3}-\frac{64 (\gamma -2) V ^ {(1)0}(w)\,\pa V ^ {(1)0}(w) a_{(2)++}^2}{3 k_{\rm cs} (w-z)} \\
& & \nn -\frac{16 (\gamma -3) (\gamma -2) G^{'-+}(w)\,G^{'+-}(w) a_{(2)++}^2}{3 k_{\rm cs} (w-z)}\\
\nn  &&-\frac{64 (\gamma -2)^2 G^{--}(w)\,G^{++}(w) a_{(2)++}^2}{9 k_{\rm cs} (w-z)}-\frac{128 (\gamma -2) \gamma  A^{+3}(w)\,V ^ {(1)0}(w)\,V ^ {(1)0}(w) a_{(2)++}^2}{3 k_{\rm cs}^2 (w-z)}\\
\nn  &&-\frac{256 (\gamma -2) (\gamma -1) A^{--}(w)\,A^{-+}(w) a_{(2)++}^2}{3 k_{\rm cs} (w-z)^2}-\frac{256 (\gamma -2) (\gamma -1) A^{-3}(w)\,A^{-3}(w) a_{(2)++}^2}{3 k_{\rm cs} (w-z)^2}\\
 \nn &&-\frac{128 (\gamma -2) \gamma  (3 \gamma -2) A^{+-}(w)\,A^{++}(w) a_{(2)++}^2}{3 k_{\rm cs} (w-z)^2}-\frac{256 (\gamma -2) \gamma  (3 \gamma -1) A^{+3}(w)\,A^{+3}(w) a_{(2)++}^2}{3 k_{\rm cs} (w-z)^2}\\
\nn  &&-\frac{512 (\gamma -2) \gamma  A^{+3}(w)\,V ^ {(2)0}(w) a_{(2)++}^2}{3 k_{\rm cs} (w-z)a_{(2)k}}
-\frac{128 (\gamma -2) a_{(2)++}^2 \big(2a_{(2)k} T(w)-2 V ^ {(2)0}(w)+3 \gamma a_{(2)k} \pa A^{+3}(w)\big)}{3 (w-z)^2a_{(2)k}}\\
\nn  &&+\frac{8 a_{(2)++}^2 }{3 (w-z)a_{(2)k}}\big(16a_{(2)k} \pa^2 A^{+3}(w) \gamma ^2+16a_{(2)k} T'(w) \gamma -16 \pa V ^ {(2)0}(w) \gamma \\
&&-32a_{(2)k} \pa^2A^{+3}(w) \gamma +3a_{(2)k} V^{(3)+3}(w)-32a_{(2)k} T'(w)+32 \pa V ^ {(2)0}(w)\big)
-\frac{128 k_{\rm cs}(\g-2)a^2_{(2)++}}{(z-w)^4} \ ,\label{VppVpm}
\end{eqnarray}}
and
{\footnotesize
\begin{eqnarray}
\nn V^{(2) 0}(z)V^{(2) 0}(w) &\sim&
\frac{(\gamma -1) (\gamma +1)a_{(2)k}^2 A^{--}(w)\,\pa A^{-+}(w)}{(\gamma -2) k_{\rm cs} (w-z)}
+ \frac{(\gamma -1) (\gamma +1)a_{(2)k}^2 A^{-+}(w)\,\pa A^{--}(w)}{(\gamma -2) k_{\rm cs} (w-z)} \nn \\
\nn  &&+\frac{2 (\gamma -1) (\gamma +1)a_{(2)k}^2 A^{-3}(w)\,\pa A^{-3}(w)}{(\gamma -2) k_{\rm cs} (w-z)}
-\frac{\gamma  (\gamma +1)a_{(2)k}^2 A^{+-}(w)\,\pa A^{++}(w)}{(\gamma -2) k_{\rm cs} (w-z)} \nn \\
\nn  &&-\frac{\gamma  (\gamma +1)a_{(2)k}^2 A^{++}(w)\,\pa A^{+-}(w)}{(\gamma -2) k_{\rm cs} (w-z)}-\frac{2 \gamma  (\gamma +1)a_{(2)k}^2 A^{+3}(w)\,\pa A^{+3}(w)}{(\gamma -2) k_{\rm cs} (w-z)}\\
\nn  &&-\frac{(\gamma +1)a_{(2)k}^2 V ^ {(1)0}(w)\,\pa V ^ {(1)0}(w)}{2 (\gamma -2) k_{\rm cs} (w-z)}+\frac{2 \gamma  (\gamma +1)a_{(2)k}^2 A^{+-}(w)\,A^{++}(w)}{(\gamma -2) k_{\rm cs} (w-z)^2}\\
\nn  &&+\frac{2 \gamma  (\gamma +1)a_{(2)k}^2 A^{+3}(w)\,A^{+3}(w)}{(\gamma -2) k_{\rm cs} (w-z)^2}+\frac{(\gamma +1)a_{(2)k}^2 V ^ {(1)0}(w)\,V ^ {(1)0}(w)}{2 (\gamma -2) k_{\rm cs} (w-z)^2}\\
\nn  &&-\frac{2 (\gamma -1) (\gamma +1)a_{(2)k}^2 A^{--}(w)\,A^{-+}(w)}{(\gamma -2) k_{\rm cs} (w-z)^2}
-\frac{2 (\gamma -1) (\gamma +1)a_{(2)k}^2 A^{-3}(w)\,A^{-3}(w)}{(\gamma -2) k_{\rm cs} (w-z)^2}\nn \\
 &&+\frac{a_{(2)k} \big(\gamma a_{(2)k} T'(w)+a_{(2)k} T'(w)-2 \gamma  \pa V ^ {(2)0}(w)+\pa V ^ {(2)0}(w)\big)}{(\gamma -2) (w-z)}
\label{v2kv2k} \\
\nn  &&-\frac{2a_{(2)k} (\gamma a_{(2)k} T(w)+a_{(2)k} T(w)-2 \gamma  V ^ {(2)0}(w)+V ^ {(2)0}(w))}{(\gamma -2) (w-z)^2}
-\frac{3 (\gamma +1) k_{\rm cs} a_{(2)k}^2}{(\gamma -2) (w-z)^4}\ .
\end{eqnarray}}
The remaining OPEs we have worked out can be found in the ancillary file on the {\tt arXiv}.

\section{The truncation analysis for the linear $A_\g$ algebra}\label{agammanull}

In this appendix we repeat the truncation analysis of section~\ref{wtruncation} for the linear $A_\g$ algebra.
The ansatz for the null-vectors is now
\bea
\cn^+  & = &
\bigl(G^{+-}_{-1/2}G^{++}_{-1/2}
+a\, Q^{+-}_{-1/2}Q^{+-}_{-1/2}
+b\, Q^{+-}_{-1/2}G^{+-}_{-1/2}
+c\, G^{+-}_{-1/2}Q^{+-}_{-1/2}
+d\, A^{++}_{-1}\bigr)\F_s \nn \\[4pt]
\cn^- &= & \bigl(G^{++}_{-1/2}G^{-+}_{-1/2}
+e\, Q^{++}_{-1/2}Q^{-+}_{-1/2}
+f\, Q^{++}_{-1/2}G^{-+}_{-1/2}
+g\, G^{++}_{-1/2}Q^{-+}_{-1/2}
+h\,  A^{-+}_{-1}\bigr)\F_s\ , \nn
\eea
where the $Q$'s are the modes of the free fermions, see \cite{Gaberdiel:2013vva}  for our conventions. Here
$\F_s$ is again a state that corresponds to the bottom component of the $R^{(s)}$ multiplet.
These vectors are null provided that they are annihilated by the positive modes of the $A_\g$ generators.
Again, there are two sets of solutions for  $\cn^+$:
\bea
\nn \text{solution 1}:&& a= \frac{-4 s}{k^-+2}\,,\quad b= \frac{2}{k^-+2}\,,\quad c= \frac{2}{k^-+2}\,, \quad d= \frac{4 (k^-+2) s+4}{k^-+2}\,,\quad k^+= 2\\[4pt]
\nn \text{solution 2}:&& a= \frac{4 (s+1)^2}{(k^-+1)^2}\,,\quad b= \frac{2 (s+1)}{k^-+1}\,,\quad c= \frac{2 (s+1)}{k^-+1}\,,\quad d= 0\,,\quad k^+= \frac{1-k^- s}{s+1} \ .
\eea
These solutions correspond to (\ref{nullm}) under the replacement of $k^\pm \mapsto k^\pm -1$.
Similarly, $\cn^-$ becomes null for
\bea
\nn \text{solution 1}:&& \hspace{-5mm}
e = \frac{-4 s}{k^++2}\,,\quad
f= \frac{-2}{k^++2}\,,\quad
g= \frac{-4 ((k^++2) s+1)}{k^++2}\,,\quad
h= \frac{-2}{k^++2}\,,\quad k^-= 2\\[4pt]
\nn \text{solution 2}:&& \hspace{-5mm}
e= \frac{4 (s+1)^2}{(k^++1)^2}\,,\quad
f= -\frac{2 (s+1)}{k^++1}\,,\quad
g= 0\,,\quad
h= -\frac{2 (s+1)}{k^++1}\,,\quad k^-= \frac{1-k^+ s}{s+1}\ ,
\eea
which correspond, upon setting $k^\pm \mapsto k^\pm -1$, to the solutions found in (\ref{nullp}).
Both vectors become simultaneously null for  $k^+=k^-=2$ or
\be
k^+=k^-=\frac {1} {1 + 2 s}=1+\frac {-2s} {1 + 2 s}\ .
\ee

\end{document}